\documentclass[a4paper,11pt]{article} 
\pdfoutput=1
\usepackage{subfig}
\usepackage{jheppub}
\usepackage{microtype}
\usepackage{lmodern}
\usepackage[T1]{fontenc}
\usepackage{amssymb}
\usepackage{mathptmx}
\usepackage{graphicx}
\usepackage[normalem]{ulem}
\usepackage{color}
\let\normalcolor\relax 
\definecolor{myred}{rgb}{0.6,0,0} 
\definecolor{myblue}{rgb}{0,0.2,0.4}
\definecolor{mygreen}{rgb}{0,0.9,0.1}
\definecolor{hc}{rgb}{.9,0.1,0.7}
\definecolor{hcout}{rgb}{.9,0.7,0.9}
\definecolor{Orange}{rgb}{1.,0.65,0.}
\usepackage{amsmath}
\usepackage{rotating}
\usepackage{textcomp}
\usepackage{timestamp} 

\numberwithin{equation}{section}
\numberwithin{figure}{section}
\numberwithin{table}{section}

\newcommand{\be}{\begin{equation}}
\newcommand{\ee}{\end{equation}}
\newcommand{\bea}{\begin{eqnarray}}
\newcommand{\eea}{\end{eqnarray}}
\newcommand{\non}{\nonumber}

\newcommand{\ta}[1]{#1\hspace{-.42em}/\hspace{-.07em}} 

\title{\boldmath
Complete QED NLO contributions to the reaction $e^+e^- \to \mu^+\mu^-\gamma$
 and their implementation in the event generator PHOKHARA\footnote{Dedicated to the memory of our late collegue and friend Jochem Fleischer (1937 - 2013).}
}

\preprint{FTUV-13-0212 \;\; IFIC/13-87\;\;\\{\hspace*{10cm} LPN13-094\;\;SFB/CPP-13-108}}

\author[a]{F.~Campanario,}
\author[b]{H.~Czy\.z,}
\author[b]{J.~Gluza,}
\author[b]{M.~Gunia,}
\author[c]{T.~Riemann,}
\author[a]{G.~Rodrigo}
\author[d]{and V.~Yundin}

\affiliation[a]{Instituto de F\'isica Corpuscular, Universitat de Val\'encia -
 Consejo Superior de Investigaciones
Cient\'ificas, \\ Parc Cient\'ific, 46980 Paterna, Valencia, Spain.}
\affiliation[b]{ Department of Field Theory and Particle Physics,
      Institute of Physics, \\
      University of Silesia, Uniwersytecka 4, PL-40-007 Katowice,
      Poland}
\affiliation[c]{Deutsches Elektronen-Synchrotron, DESY, Platanenallee
    6, 15738 Zeuthen, Germany}
\affiliation[d]{Max-Planck-Institut f\"ur Physik, F\"ohringer Ring 6, 80805 Munich, Germany}
\emailAdd{francisco.campanario@ific.uv.es}
\emailAdd{czyz@us.edu.pl}
\emailAdd{janusz.gluza@us.edu.pl}
\emailAdd{guniamichal@gmail.com}
\emailAdd{tord.riemann@desy.de}
\emailAdd{german.rodrigo@csic.es}
\emailAdd{yundin@mpp.mpg.de}

\abstract{KLOE and Babar have an observed discrepancy of 2\% to 5\%  in the invariant pion pair
production cross section. These measurements are based on approximate
NLO $\mu^+ \mu^-\gamma$ cross section predictions of the Monte Carlo event generator PHOKHARA7.0.
In this article, the complete NLO radiative corrections to $\mu^+ \mu^- \gamma$
production are calculated and implemented in the Monte Carlo event generator PHOKHARA9.0.
Numerical reliability is guaranteed by two independent approaches 
to the real and the virtual corrections.
The novel features include the contribution of pentagon diagrams in the virtual
corrections, which form a gauge-invariant
set when combined  with their box diagram partners.
They may contribute to certain distributions at the percent level.
Also the real emission was complemented with two-photon final state
emission contributions not included in the generator PHOKHARA7.0.
We demonstrate that the numerical influence reaches, for realistic charge-averaged experimental setups, not
more than 0.1\% at KLOE and 0.3\% at BaBar energies.
As a result, we exclude the approximations in earlier versions of PHOKHARA as origin of the observed
experimental discrepancy.
}


\begin{document}

\maketitle


\section{Introduction \label{sec-intro}}
The total cross section for
electron-positron annihilation into hadrons is a very important
physical observable weighting significantly on the theory error
of the muon anomalous magnetic moment and the running electromagnetic
 fine structure constant used in the tests of the Standard Model and its extensions.
For recent reviews see e.g.~\cite{Actis:2010gg,Hagiwara:2011af,Davier:2010nc,Jegerlehner:2009ry,Harlander:2002ur}.
 At high energies, the cross section can be calculated using perturbative QCD,
 however at low energies one has to rely on the experimental measurements.

One of the methods used to extract the hadronic cross section
is the
''radiative return'', exploiting the fact that the cross section
of the process with initial state photon radiation can be factorised
into a known perturbative factor and the hadronic cross section
without initial state radiation at the energy lowered by the emitted
  photons.  Due to the complexity of the experimental setup, the extraction of the hadronic cross
 section within a realistic
 experimental framework can be achieved only by means of an event generator. 

 The most important contribution to the hadronic cross section
 is the pion pair production channel. Its accuracy is an issue 
 as it provides the main source of error in the evaluation of the muon anomalous 
 magnetic moment~\cite{Hagiwara:2011af}. In view of the planned improvement of the direct measurement
 of the muon anomalous magnetic moment~\cite{Venanzoni:2012vha}, with the expected error four times smaller than
 the present one, it is crucial to pull down the theoretical error as much as possible. 
 The pion production cross section in $e^+e^-$
 scattering was measured, using the radiative return method, by BaBar~\cite{Lees:2012cj,Aubert:2009ad} and
KLOE~\cite{Babusci:2012rp,Ambrosino:2010bv,Ambrosino:2008aa}.
 Both experiments quote individual errors at the level of a fraction of a percent, while the discrepancy
 between them is up to 2\% at the $\rho$ peak and 5\% when approaching the energy of 1~GeV. 
The origin of the discrepancy remains unclear.
 Since both experiments use the PHOKHARA~\cite{Czyz:2002np,Czyz:2005as}
 event generator to extract the hadronic cross section, it is necessary to
 check carefully its physical content. The PHOKHARA event generator was used to generate the 
 reactions $e^+e^-\to \pi^+\pi^- + \ {\rm photons}$ and $e^+e^-\to \mu^+\mu^- + \ {\rm photons}$. 
 The
 latter process is used for monitoring the luminosity. So far, the version of PHOKHARA used
 by BaBar and KLOE included the dominant next-to-leading order (NLO) radiative corrections. 
In view of the above mentioned  discrepancy between BaBar and KLOE, it is essential to make a full NLO
calculation and to establish the importance of the missing contributions.

In this article, the 
complete radiative NLO Quantum Electrodynamics~(QED) corrections to the reaction
  $e^+e^-\to \mu^+\mu^- \gamma$ are calculated, tested and implemented into the event generator 
 PHOKHARA.
From a technical point of view, the pentagon diagrams are the most challenging.
Because there is no scale available which might lead to logarithmically enhanced contributions, they are
expected to be small. However, it is known that logarithmic enhancements
  can be
 generated in some regions of the phase space within
 complicated experimental setups. 
The fact that we can not neglect the small electron mass for the same reason poses an additional
 challenge in the calculation of the virtual amplitudes since ratios of the order of $s/m_e^2$ can appear, where $s$ is the
 energy of the collider. This demands a good control on the numerical
 accuracy of our amplitudes~\footnote{Technically, the accuracy problems in the calculation
of the radiative corrections are similar to the ones in
$e^+e^-\to \bar t t \gamma$,  solved in Ref.~\cite{Khiem:2012bp}
using the GRACE system.  The electroweak radiative corrections were calculated there, but with 
 the photon emitted at large angles only.}.
Consequently, detailed Monte Carlo studies are mandatory. 
First studies were presented some time ago by the Ka\-to\-wi\-ce/Zeuthen group~\cite{Kajda:2009aa}, in PhD
theses~\cite{kajphd,yundin-phd-2012--oai:export} and also by an independent
alternative approach~\cite{Actis:2009zz,Actis:2009uq}. 
This enabled us to compare specific phase space points with~\cite{Actis:2009zz,Actis:2009uq} with high
precision as a first numerical test; see
Ref.~\cite{yundin-phd-2012--oai:export} for details.
Here, we perform a complete calculation in the frame of a realistic Monte Carlo environment, PHOKHARA9.0.
Because there are several known sources of numerical instabilities, and because we have no external
cross-check available, we organized for two independent implementations of the QED virtual corrections. 
Having implemented the complete radiative corrections, detailed physics studies
 became possible, and their results are presented here.

 The article is organised as follows: In Section~\ref{sec:class} and Appendices~\ref{appex:twophoton} and~\ref{sec-asoft} we give a detailed description
 of the calculation of the radiative corrections. Section~\ref{sec-impl}
 sketches the implementation of the radiative corrections into 
 PHOKHARA9.0. 
In Section~\ref{sec-tests}, 
 the main tests of the correctness and the numerical stability  of the code are presented.
Further, the relevance of the NLO
radiative corrections, which were missing 
  in PHOKHARA7.0, is investigated.
In Section~\ref{sec-impact}, the 
possible impact of the radiative corrections, analogous to those studied here, on the  pion
form factor measurements of BaBar and KLOE is discussed. 
 Section~\ref{sec-conc} contains the conclusions.

 \section{Radiative corrections to $e^+ e^- \to \mu^+ \mu^- \gamma$ \label{sec:class} }
 
\begin{figure}[htb]%
    \centering
     \includegraphics[width=1.\textwidth]{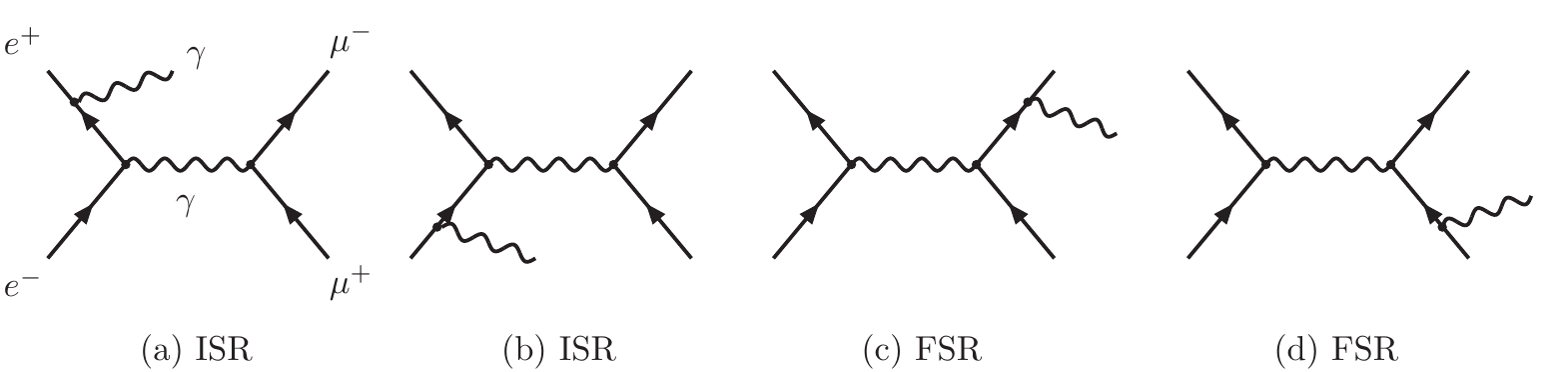} %
    \caption{The tree diagrams for $e^+ e^- \to \mu^+ \mu^- \gamma$.} %
    \label{fig:eemmg0}
\end{figure}

The tree level diagrams contributing to the leading order (LO) amplitude 
are shown in Fig.~\ref{fig:eemmg0}.  
There are two types of contributions, those  with initial state photon emission (ISR)
and final state photon emission (FSR). 
The ISR and FSR pairs of diagrams are separately gauge invariant.  

At NLO QED,   
there are the virtual and the real corrections resulting in three
types of contributions to the cross section, the ISR and the FSR contributions, and their ISR-FSR interference terms.

We use dimensional regularization~\cite{'tHooft:1972fi} to regularize the ultraviolet (UV) and 
infrared (IR) divergences. 
The UV divergences of the virtual amplitude are removed by the renormalization
counter-terms.
Both the virtual and the real corrections are infrared divergent. These divergences cancel
in the sum for infrared-safe observables. The IR divergences are canceled 
and both the virtual+real~(soft) and the real~(hard) corrections
 become separately numerically integrable.  Details of
the real emission calculation are given in
Section~\ref{sec-nlo-real}. In the following, we describe the method used to
compute the virtual amplitudes.

 \subsection{Virtual corrections \label{sec:class1}  }

\begin{figure}[htb]%
    \centering
     \includegraphics[width=1.\textwidth]{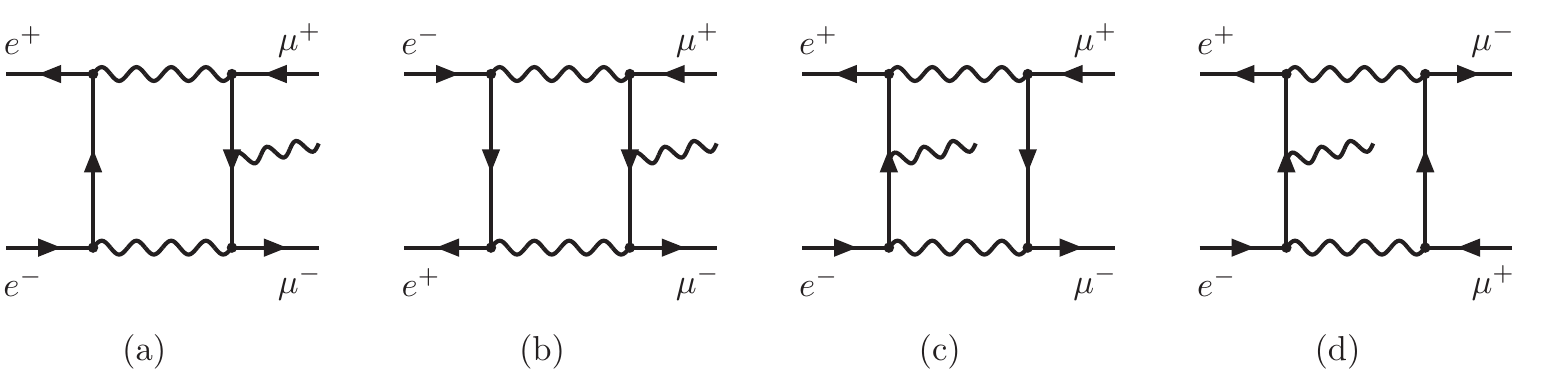} %
    \caption{The four diagrams with  pentagon topology for $e^+ e^- \to \mu^+
      \mu^- \gamma$} %
    \label{fig:pentagons}
\end{figure}

Besides photonic self-energy corrections, there are 32 diagrams contributing to $e^+e^-\to\mu^+ \mu^-\gamma$
at NLO QED. They can be classified into several independent gauge invariant
subsets, which we will call Penta-Box, Box-Triangle-Bubble and Triangle
contributions. The first class involves loop corrections with the two
lepton lines attached to the loop. The most challenging diagrams are the four pentagon diagrams,
shown in  Fig.~\ref{fig:pentagons}, 
where a real photon is emitted from an  internal line.

\begin{figure}[h!]%
    \centering
   \includegraphics[width=1.\textwidth]{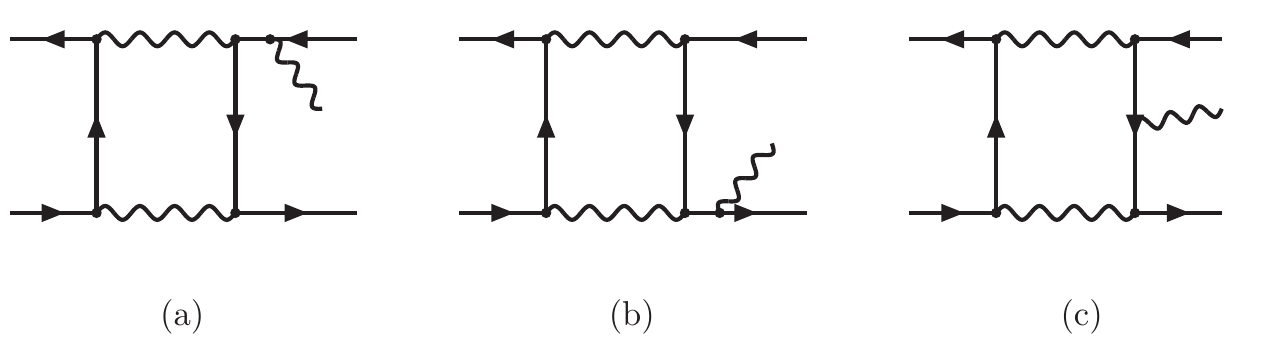} %
    \caption{One of the four gauge invariant combinations of a pentagon with two boxes with external photon
emission for $e^+ e^- \to \mu^+ \mu^- \gamma$.  }%
    \label{fig:pentabox}
\end{figure}

They do not constitute a class of gauge independent diagrams by themselves.
Gauge invariant groups are formed when a pentagon is associated with
two box diagrams where a photon is radiated from the same external (electron or muon) line. This is shown
schematically in Fig.~\ref{fig:pentabox}.
The contribution of these twelve Penta-Box diagram combinations, interfering with the tree level
diagrams of Fig.~\ref{fig:eemmg0}, will be discussed in detail in Section~\ref{sec-tests}.

\begin{figure}[h!]%
    \centering
   \includegraphics[width=1.\textwidth]{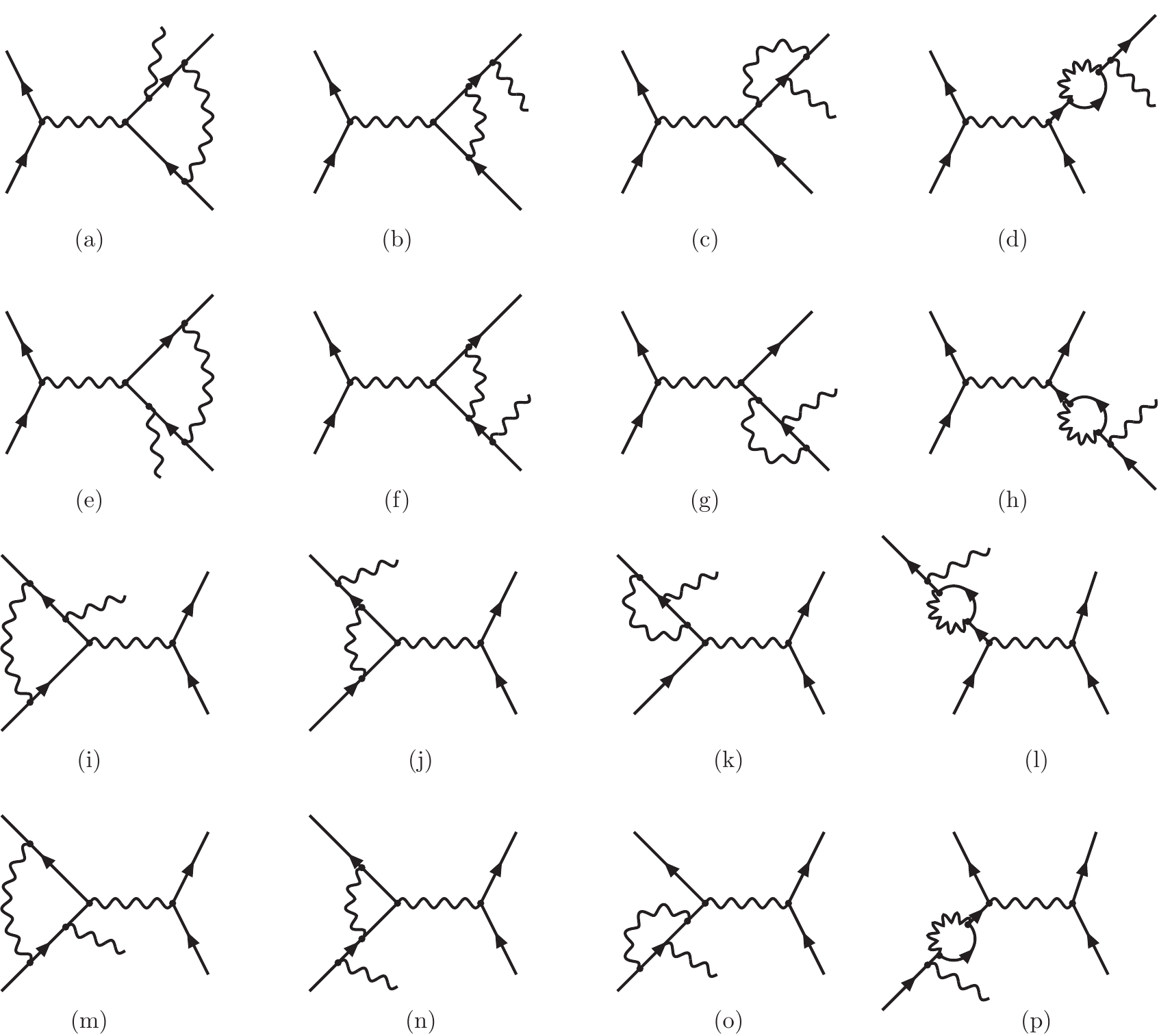} %
    \caption{The  set of sixteen one-loop Box, triangle  and self-energy diagrams with internal photon emission  in $e^+ e^-
\to \mu^+ \mu^- \gamma$.  
}%
    \label{fig:wholeBT}
\end{figure}

\begin{figure}[h!]%
   \centering
   \includegraphics[width=1.\textwidth]{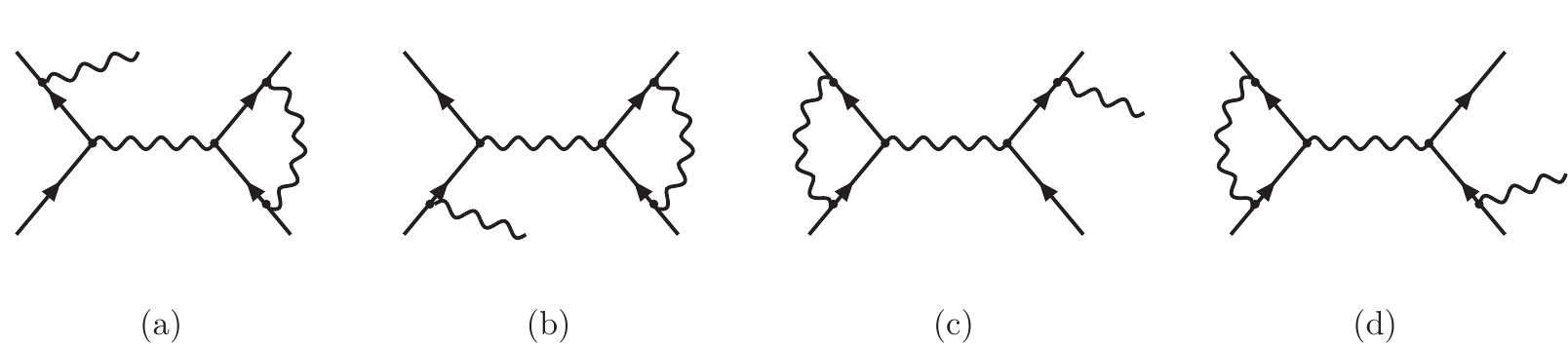} %
    \caption{The four triangle diagrams with external photon emission  in $e^+ e^- \to \mu^+ \mu^- \gamma$.}%
    \label{fig:triangle}
\end{figure}

Box-Triangle-Bubble and Triangle
contributions contain corrections to one lepton line (electron or muon) and are further
classified depending whether the loop and the real photon are attached to the
same lepton line.
The Box-Triangle-Bubble class contains all the loop corrections to a lepton 
line with a real (on-shell) photon and a second off-shell photon, connecting to
the other lepton line. The contributing  boxes, vertices and bubbles can be found in
Fig.~\ref{fig:wholeBT}. There are two independent gauge invariant subsets, for
FSR (two upper lines of Fig.~\ref{fig:wholeBT}) and for ISR~(two lower lines of Fig.~\ref{fig:wholeBT}).
The triangle
contributions are given in Fig.~\ref{fig:triangle}. There, a real photon is emitted from one fermion line, 
and the other photon (off-shell) entering a 3-point function is connected to the
other fermion line.  

Finally, we mention the diagrams with external photon emission and self-energy insertions to the photon
propagator~\footnote{The sum of contributions
from diagrams with real emission from fermionic self-energy insertions to the photon propagator vanishes due
to the Furry theorem.}.
They constitute a gauge-invariant universal correction which can be accounted for in any QED calculation by
simply
running the fine structure constant to the appropriate scale~\cite{Bohm:1986rj,Denner:1991kt}.
These self-energies  are treated separately and have been omitted  from our fixed-order loop amplitude
definition in Fig.~\ref{fig:wholeBT}.
The treatment of vacuum polarisation in the PHOKHARA event generator, together 
 with narrow resonance contributions is described in detail 
in Ref.~\cite{Czyz:2010hj} and will not be discussed here.

In the present article, two independent programs using two different  methods are used.  One is based on
the trace method and the calculations are done using double precision
numerical routines, including the \texttt{PJFry} libraries~\cite{pjfry-project}. 
We refer to it as ``Double precision - Trace method'' (DT-method). 
The other one is based on the helicity formalism as  
  described in Ref.~\cite{Campanario:2011cs}, and we refer to it as ``Quadruple precision - Helicity method''
(QH-method) because numerical calculations are done partially using
quadruple  precision. Such independent implementations are necessary to gain
sufficient numerical reliability.

\subsubsection{The DT-method \label{subsubsec-DT}}

With the DT-method, topologies are generated by QGRAF~\cite{Nogueira:1991ex} and then dressed with
particles and
momenta
by the DIANA program~\cite{Tentyukov:1999is} according to the QED model description file.
The resulting output contains a list of Feynman diagrams in the textual representation,
which is defined by the TML markup language script~\cite{Tentyukov:1999yq}.
Next, the diagrams are passed through the FORM~\cite{Vermaseren:2000nd} script, 
which substitutes Feynman rules according to the selected model.
Further manipulations are done with FORM.
In addition, some general simplifications can be enabled by setting configuration parameters.
This includes gamma algebra identities like $\gamma^\mu\gamma^\nu\gamma_\mu=(2-d)\gamma^\nu$,
the transversality condition, e.g.  $p_1 \cdot \epsilon(p_1)=0$, usage of Dirac equation and momentum
conservation.
 The resulting expressions are written in the FORM tablebase.
We use it as an input in the squaring program which sums the diagrams and multiplies them
by the complex conjugated set of Born diagrams. 
The fermion lines are connected by the completeness relation. Then, Dirac traces are taken.

For the calculation of the newly added one-loop pentagon contributions, 
one has to calculate 5-point tensor Feynman integrals up to rank $R=3$. 
We reduce the tensor integrals in $d=4-2\epsilon$ dimensions to scalar 1-
to 4-point functions.
They depend on the reduction basis chosen.
Often one uses as basis momenta, the external momenta of the diagram as in Refs.~\cite{Passarino:1978jh,Campanario:2011cs}.
Our choice (with a one-to-one correspondence) are the so-called chords, the shifts of internal momenta
with respect to the loop momentum~\cite{pjfry-project}. 
 
Advanced tensor integral calculations became a standard task in recent years,
mainly
triggered by LHC physics.
Nevertheless, ensuring sufficient numerical stability is  demanding for several reasons.
An often discussed issue is the treatment (or avoidance) of small or vanishing inverse Gram determinants.
Another one is just the extreme spread of scales met in our physical process, because we cannot 
neglect
the electron mass $m_e \approx 1/2000$ GeV as an
independent parameter.
With $\sqrt{s} = 1 - 10$ GeV, one faces e.g. a ratio $m_e^2/s \sim 10^{-7}-10^{-9}$.  
%
The DT implementation of tensor integral calculation
relies on the approach developed in Refs.~
\cite{Davydychev:1991va,Tarasov:1996br,Fleischer:1999hq,Diakonidis:2008ij,Diakonidis:2009fx,
Fleischer:2010sq,Almasy:2013uwa}
and uses the PJFry tensor reduction package~\cite{yundin-phd-2012--oai:export,Fleischer:2011zz},
combined with QCDLoop/FF~\cite{Ellis:2007qk,vanOldenborgh:1990yc} or OneLOop~\cite{vanHameren:2010cp} for
scalar integrals.
More technical details can be found in Ref.~\cite{yundin-phd-2012--oai:export}~\footnote{A new approach to the treatment of pentagon diagrams is under development in the OLEC project~\cite{olec-project-2013,Fleischer:2012ad,Almasy:2013uwa}.
It is alternative to tensor
reduction and relies instead on the direct calculation of tensor contractions~\cite{Fleischer:2010sq,Fleischer:2011nt}.
It will be interesting to see whether this improves speed or stability of the numerics.}.

\subsubsection{The QH-method \label{subsubsec-QH}}

\begin{figure}[h!]%
    \centering
   \includegraphics[width=1\textwidth]{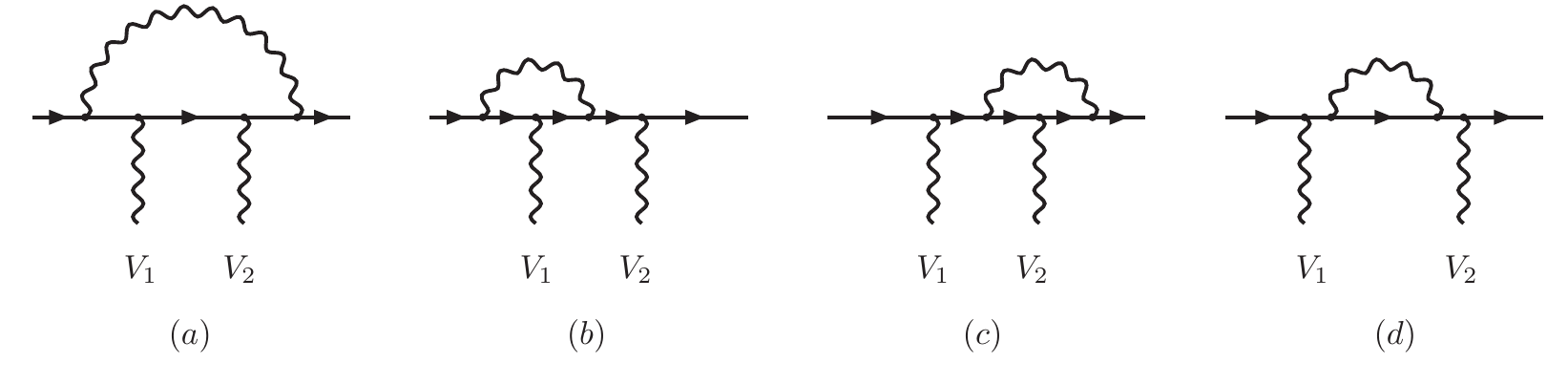} %
    \caption{Boxline contributions for $e^+ e^- \to \mu^+ \mu^- \gamma$. }%
    \label{fig:boxtriangle}
\end{figure}

The second implementation (QH-method) uses the  helicity formalism as  
  described in Ref.~\cite{Campanario:2011cs}. To build the virtual amplitude, four
  building blocks are used. Corrections to a lepton line with two real
  (on-shell or off-shell) photons 
  attached in a fixed order of external
  bosons, Fig.~\ref{fig:boxtriangle}, constitute the first building block, which we call
  Boxline and also include the corresponding counter-terms which are not shown in Fig.~\ref{fig:boxtriangle}. We used the effective current approach, thus, $V_1$ and $V_2$ should be understood as generic off-shell
currents which can be, in this case, an on-shell photon or an off-shell
photon, which forms the second lepton line.  The  physical amplitude is built by considering
  all physical permutations and contractions with external
  currents yielding the Box-Triangle-Bubble gauge invariant
  subsets of Fig.~\ref{fig:wholeBT}. In addition, we use the
  vertex corrections to a lepton line with one real photon attached to it. All
  possible contributions result in the triangle contributions of
  Fig.~\ref{fig:triangle}. %
The third
building group is formed by the Penta-Box diagrams depicted in
Fig.~\ref{fig:pentabox}, which involve the
pentagon diagrams. The last building block is obtained by crossing the two
initial fermion lines in Fig.~\ref{fig:pentabox} and constitute an
independent gauge group.
To compute them, we generalized the software developed
in Ref.~\cite{Campanario:2011cs} to be able to compute diagrams with two
fermion lines. This includes the use of  Chisholm identities~\cite{Sirlin:1981pi} which reduces
the CPU time required to evaluate the Penta-Box contributions by a factor
ten. At the same time, this improves the stability of the code since it makes
explicit terms proportional to the small electron mass. %
The calculation of tensor 
integrals is done by using Passarino-Veltman reduction~\cite{Passarino:1978jh} 
for up to $4$-point diagrams and the method of~\cite{Denner:2005nn,Binoth:2005ff}, following
the convention of Ref.~\cite{Campanario:2011cs} for higher-point tensor integrals. 
The scalar integrals are calculated as in
Refs.~\cite{'tHooft:1978xw,Denner:1991kt}. %

We use a cache system in all the building blocks such that the
information of the loop-dependent parts are stored. This is particularly important for this process since up to 32
different helicity amplitudes exist corresponding to the different helicity
and polarization combinations of the external particles. After the first
helicity is computed, which include the evaluation of the loop-dependent parts, any additional helicity amplitude is computed with less than
10$\%$ of CPU time, reducing the CPU time of the code by a factor 10.

The building blocks do not use special properties like 
transversality or being on-shell for the real photons attached to the lepton
lines, instead, we assume external effective current attached to them. This
allows us to use Ward identities, by replacing
an effective current with the corresponding momentum, to check the accuracy of the computed amplitudes. We classified our
contributions in gauge invariant subsets so that the Ward
identities are fulfilled.  Those identities are called gauge tests and are
checked with a small additional computing cost, using the
cache system. They are checked for every phase space point and each gauge invariant subset distinguishing between
FSR and ISR contributions.
This is important because the phase space integration 
of the virtual contribution shows numerical instabilities 
in the calculation of the one-loop tensor integrals~\cite{Campanario:2011cs}.

We have implemented a rescue system for
phase space points where the Ward identities are not satisfied with an
accuracy of at least three digits.  First, we calculate the amplitudes applying quadruple precision only
to the scalar integrals and tensor reduction routines. This requires reconstructing the external momenta
in quadruple precision, so that global energy-momentum conservation is still fulfilled at the higher numerical accuracy retaining the external particles on their mass-shell.
If the Ward identities are not satisfied, the rescue system evaluates the 
amplitude using quadruple precision in all parts of the code.
With this system, we find that the proportion of phase space points that do not pass 
the Ward identities for a requested accuracy of
$\epsilon=10^{-3}$ is well below one in ten thousand. The rescue system
adds an additional $10\%$ to the CPU time.

Despite the cache system and the use of Chisholm identities, this implementation is
still seven times slower that the DT-method which can be traced back to the
evaluation of the 32 helicity amplitudes and the parts evaluated in quadruple precision.

 We have tried to evaluate the
amplitudes only in double precision to improve the speed of the code by a
factor two using dedicated
subroutines for small Gram determinants. These involve the evaluation of
three- and four-point functions up to Rank 11 and 9, respectively,
following the notation of Ref.~\cite{Campanario:2011cs}.  The high rank of the
rescue system allows to obtain full double precision for mild cancellations
in the Gram determinants.  However, two problems arise. First, for the failing phase space points, there was
always some internal combination for which the expansion breaks down due to the
presence of additional cancellations in sub-Cayley determinants. Thus, 
full double precision is not achieved for all tensor
integrals coefficients. Second, within the
helicity method formalism, there exist extremely large cancellations between the different helicity
amplitudes resulting into an additional loss of precision which can be larger
than the one due to the presence of small Gram determinants. These numerical cancellations are related to 
the fact that the mass of the electron has to be retained, and many numerical
cancellation occur. For example, in the numerical calculation of 
$\slash \kern-6pt {p}_e u(p_e)= m_e u(p_e) $ cancellations of the order of
 $s/m_e^2$ would appear, where s is the energy
of the collider, if the Dirac equation is not applied or is not treated
carefully.

These two problems are reflected in a bad accuracy of the gauge test and,
therefore, in the large number of identified unstable points using
only double precision. The second problem is naturally solved using the DT-method since the
summation and averaging over spins are done analytically and
many of these numerical cancellations are avoided. 
We decided to implement the DT-method in PHOKHARA9.0 and compare it with the code implemented in full quadruple precision where the gauge test ensures the numerical
accuracy of the code.

\subsection{Real photon emission \label{sec-nlo-real}}

 The real two-photon emission, which contributes to the $e^+e^-\to\mu^+\mu^-\gamma$
cross section at NLO is now included in the PHOKHARA code
 completely in contrast to the implementation of Refs.~\cite{Rodrigo:2001kf,Czyz:2004rj}, where
 subleading contributions were neglected.  We distinguish between soft and 2-hard photon
 emission. 
\subsubsection{Soft photon emission \label{sec-soft}}
  
  In the soft photon contribution, the phase space of one of the photons ($k_1)$
  is integrated out analytically. The integrals to be performed, which factorise
 in front of the square of the full amplitude describing the
 $e^+e^-\to\mu^+\mu^-\gamma$~ reaction, read
\begin{eqnarray}
F(p_1,p_2,q_1,q_2,r) = \frac{-\alpha}{4 \pi} \int \frac{d^3k_{1}}{E_{k_1}} \left[ \left( \frac{p_1}{p_1 \cdotp  k_1} - \frac{p_2}{k_1 \cdotp p_2} \right) +  \left( \frac{q_2 }{q_2 \cdotp  k_1} - \frac{q_1}{k_1 \cdotp q_1} \right) \right]^2,\label{ffdef}
\end{eqnarray}
where $p_1$, $p_2$, $q_1$ and  $q_2$  are the momenta of the positron,
electron, antimuon and muon, respectively.
 The infrared regulator --- the photon mass $\lambda$ and the photon energy
 cut-off
$E_{max}$ dependence can be cast into a single parameter
\begin{eqnarray}
r=\frac{2E_{max}}{\lambda}. 
\end{eqnarray}

 In principle, this is a well known formula, which can be found in
the literature~\cite{Berends:1987ab,Rodrigo:1999qg}. However, as the 
integral over the photon energy is performed only
 up to a given cut-off $E_{max}$,
 its form depends on the frame
 in which this cut-off is applied. We found it suitable to give the formulae,
 which are valid in any frame in which the cut-off is defined. They are a bit
 longer than the usual ones as we express everything through the four momenta
 $p_1,p_2,q_1,q_2$ given in the frame, where the cut-off is defined, in contrast
  to the usual expressions which use invariants, but are less universal.
 The explicit expression for $F(p_1,p_2,q_1,q_2,r)$ is given  in
 Appendix~\ref{sec-asoft}.

\subsubsection{Two-hard photon emission \label{sec-2hard}}

For the two-hard photon emission, the helicity method for the calculation of the
amplitudes 
was used and cross checked with a dedicated code based on the trace method of spin summation. 
 The convention for the helicity amplitude method 
 introduced in Ref.~\cite{Rodrigo:2001kf} was adopted. 

The only amplitude,
  which was missing in earlier versions of PHOKHARA
 was the two-photon FSR.
 Interferences between the coded amplitudes, with infrared divergences
 matching the ones from Penta-Box diagrams,
 were also not included. 
 After some algebra similar to Refs.~\cite{Rodrigo:2001kf,Czyz:2004rj},
 a very compact form for the two-photon FSR amplitude is obtained
\begin{eqnarray}
 \kern-15pt M\left(\lambda_{\mu^+},\lambda_{\mu^-},\lambda_1,\lambda_2\right) =
 v_I^{\dagger}(\lambda_{\mu^+}) A\left(\lambda_1,\lambda_2\right)
 u_I(\lambda_{\mu^-}) 
   + v_{II}^{\dagger}(\lambda_{\mu^+}) B\left(\lambda_1,\lambda_2\right)
 u_{II}(\lambda_{\mu^-})
 \ , 
\label{eq:a1}
\end{eqnarray} 
where the matrices A and B and the convention used to define
the spinors are given in Appendix~\ref{appex:twophoton}. 
The energy of one of the photons has to be bigger than the cut-off  $E_{max}$
 and the sum of the soft and hard contributions should not depend on 
this cut-off
 up to terms $\sim E_{max}$, which are neglected in the analytic calculation.

\section{Implementation of the radiative corrections in the event generator PHOKHARA9.0
\label{sec-impl}}
PHOKHARA9.0 is available from the webpage \url{http://ific.uv.es/~rodrigo/phokhara/}.
 As stated already, all new parts of the released computer code
  were calculated independently by two methods and/or groups
 of the authors of this article.
To ensure the stability of the virtual corrections, 
 we use the two independent codes
described in Section~\ref{sec:class}.   
 The faster routine in the released version of PHOKHARA9.0 is
 used, which is the code based on the DT-method.
 The other one can be obtained on request.
 We sketch here shortly the new ingredients of the released code.
 The listed changes concern only the $e^+e^-\to \mu^+\mu^- \gamma$ mode,
 when it is running with the complete NLO radiative corrections:
 \begin{itemize}
 \item{The
  virtual corrections are calculated in 
 double precision and the sum over polarisations is done with 
 the trace method. The software PJFry~\cite{pjfry-project} is used
 for this purpose and the relevant parts of the libraries developed there
 are distributed with PHOKHARA9.0. }
 \item{The soft photon emission is calculated using the formulae discussed in 
  Section~\ref{sec-soft} and in Appendix~\ref{sec-asoft}.}
 \item{The two-hard-photon emission part uses the helicity amplitudes
  as defined in Ref.~\cite{Czyz:2004rj}. The newly added part --- the two photon FSR is described in Section~\ref{sec-2hard} and coded accordingly. }
\end{itemize}

\section{Tests of the code \label{sec-tests}}
 The released code was tested very extensively both for the real and the virtual
 contributions to assure the technical
 accuracy of the code to be much better than the one required for experimental
 measurements. The necessity of retaining a finite electron mass possesses a
 potential threat of numerical instabilities both for the real and the
 virtual contributions since cancellations of the order
 of $s/m_e^2$ can appear.

The virtual corrections constitute the most challenging part.  
The presence of Gram determinants can constitute an
additional source of instabilities which can be more challenging for some
realistic experimental selection cuts where forward photon emissions are favoured, resulting
in collinear photon emissions. 
 %

\begin{table}
\begin{center}
\begin{tabular}{|c|c|c|c|c|}
\hline
\multicolumn{5}{|c|}{$\sqrt{s}$=1.02 GeV}\\
\hline
\multicolumn{5}{|c|}{$\sigma_{QH}$ = 6.332(1) [nb] }\\
\hline
\multicolumn{5}{|c|}{$\sigma_{DT}$ = 6.332(1) [nb] }\\
\hline
$|\Delta| >$ & $\sigma_{\Delta_{DT}}$ [nb]& $\sigma_{\Delta_{QH}}$ [nb]& $\sigma_{\Delta_{DT}}/\sigma_{DT}$& N$_{event}$\\
\hline
0.1    & 0& 0             &0&0\\
\hline
0.01   & $4(4) \cdotp 10^{-8}$& $4(4) \cdotp 10^{-8}$&$6(6) \cdotp 10^{-9}$&1\\
\hline
0.001  & $1.4(3) \cdotp 10^{-6} $& $1.4(3) \cdotp 10^{-6} $&$2.2(4) \cdotp 10^{-7}$&32\\
\hline
0.0001 & $2.1(1) \cdotp 10^{-4}$& $2.1(1) \cdotp 10^{-4}$&$3.4(2) \cdotp 10^{-5}$&521\\
\hline
0.00001& $2.7(1) \cdotp 10^{-4}$& $2.7(1) \cdotp 10^{-4}$&$4.2(2) \cdotp 10^{-5}$&787\\
\hline
\end{tabular}
\end{center}
\caption{Comparison between the codes based on the DT-method  and the 
  QH-method at $\sqrt{s}$=1.02 GeV. No selection cuts are applied. See text
  for details on the definition of the Table entries. } \label{test1}
\end{table}

\begin{table}
\begin{center}
\begin{tabular}{|c|c|c|c|c|}
\hline
\multicolumn{5}{|c|}{$\sqrt{s}$=10.56 GeV}\\
\hline
\multicolumn{5}{|c|}{$\sigma_{QH}$ = 0.07004(4) [nb] }\\
\hline
\multicolumn{5}{|c|}{$\sigma_{DT}$ = 0.07004(4) [nb] }\\
\hline
$|\Delta| >$ & $\sigma_{\Delta_{DT}}$ [nb]& $\sigma_{\Delta_{QH}}$ [nb]& $\sigma_{\Delta_{DT}}/\sigma_{DT}$& N$_{event}$\\
\hline
0.1    & $6(6) \cdotp 10^{-7}$&$2(1) \cdotp 10^{-8}$  &$9(9) \cdotp 10^{-6}$&125\\
\hline
0.01   & $7(5) \cdotp 10^{-7}$&$1.4(4) \cdotp 10^{-7}$ &$1.1(9) \cdotp 10^{-5}$&1044\\
\hline
0.001  & $7.7(6) \cdotp 10^{-6} $&$7.1(2) \cdotp 10^{-6} $ &$1.10(9) \cdotp 10^{-4}$&9599\\
\hline
0.0001 & $8.3(1) \cdotp 10^{-5}$&$8.3(1) \cdotp 10^{-5}$ &$1.18(2) \cdotp 10^{-3}$&42621\\
\hline
0.00001& $2.24(2) \cdotp 10^{-4}$& $2.24(2) \cdotp 10^{-4}$&$3.21(4) \cdotp 10^{-3}$&115091\\
\hline
\end{tabular}
\end{center}
\caption{Comparison between the codes based on the DT-method  and
  QH-method at $\sqrt{s}$=10.56 GeV. No selection cuts are applied. See text
  for details on the definition of the Table entries.} \label{test2}
\end{table}

%
We performed very detailed tests for the different gauge invariant blocks
separately~\cite{Gunia:PhD}.  
Here, we mainly show the results of the tests concerning the sum 
of all contributions. 
They are summarized  in Tables~\ref{test1}-\ref{test4}. The tests were performed for two
 different energies  1.02~GeV and 10.56~GeV without any event selection
 (Tables~\ref{test1} and~\ref{test2}) and with event selections close to
 the ones used in the experiments KLOE (Table~\ref{test3}) and BaBar
 (Table~\ref{test4}) -- the specific cuts applied are found in Appendix~\ref{appex:KLOBA}.
 The integrated cross sections for both codes, DT-method ($\sigma_{DT}$)
 and QH-method ($\sigma_{QH}$), are in perfect agreement in all cases and the
 statistical errors are well below the per mille level.

For ten million of examined events with one photon in the final state,
 we count for how many events, N$_{event}$,
 the predictions for the cross section
disagree at the relative accuracy $|\Delta|$.  
%
To
 check whether these events might have an impact on the differential cross
 section, we also calculate the cross section corresponding to these events for both codes
 $\sigma_{\Delta_{DT}}$ (DT-method) and $\sigma_{\Delta_{QH}}$ (QH-method). 
At 1.02~GeV, Tabs.~\ref{test1} and~\ref{test3},  one retains at least 1 digit of accuracy for the
matrix element squared and the cross section of these events is irrelevant.
 At 10.56~GeV,  one observes in Tab.~\ref{test2} that one can lose completely
 the accuracy (the order of magnitude
  of the results is however always the same),  but that does
 not happen for the BaBar event selection cuts, Tab.~\ref{test4}, where at
 least two digits  are correct.
  Even if Table~\ref{test2} shows that the cross
 section from events for which one loses the precision is small (below 0.3\%), the
 released program based on the DT-method should be used with care at high
 energies if no event selection is applied and a cross check with the 
 QH-method is recommended. 

\begin{table}
\begin{center}
\begin{tabular}{|c|c|c|c|c|}
\hline
\multicolumn{5}{|c|}{KLOE event selection}\\
\hline
\multicolumn{5}{|c|}{$\sigma_{QH}$ = 1.575(2) [nb] }\\
\hline
\multicolumn{5}{|c|}{$\sigma_{DT}$ = 1.575(2) [nb] }\\
\hline
$|\Delta| >$ & $\sigma_{\Delta_{DT}}$ [nb]& $\sigma_{\Delta_{QH}}$ [nb]& $\sigma_{\Delta_{DT}}/\sigma_{DT}$& N$_{event}$\\
\hline
0.01  & 0&0&0&0\\
\hline
0.001  & $2(2) \cdotp 10^{-9}$&$2(2) \cdotp 10^{-9}$&$2(1) \cdotp 10^{-9}$&2\\
\hline
0.0001 & $7.7(3) \cdotp 10^{-5}$& $7.7(3) \cdotp 10^{-5}$&$4.9(2) \cdotp 10^{-5}$&713\\
\hline
0.00001& $1.02(4) \cdotp 10^{-4}$&$1.02(4) \cdotp 10^{-4}$&$6.5(2) \cdotp 10^{-5}$&1852\\
\hline
0.000001&$1.17(4) \cdotp 10^{-4}$&$1.17(4) \cdotp 10^{-4}$&$7.4(2) \cdotp 10^{-5}$&5068\\
\hline
\end{tabular}
\end{center}
\caption{Comparison between the codes based on the DT-method  and the 
  QH-method for KLOE event selection cuts~(see Appendix~\ref{appex:KLOBA}). 
The contribution $\sigma_{\Delta}$ to the cross section $\sigma$  for a 
chosen $|\Delta|$. Subscripts $QH$ and $DT$ denote
 QT-method and DT-method respectively. $q^2 \in (0.34,0.96)$ GeV$^2$. See text
  for details on the definition of the Table entries.} \label{test3}
\end{table}

\begin{table}
\begin{center}
\begin{tabular}{|c|c|c|c|c|}
\hline
\multicolumn{5}{|c|}{BaBar event selection}\\
\hline
\multicolumn{5}{|c|}{$\sigma_{QH}$ = 0.0005655(7) [nb] }\\
\hline
\multicolumn{5}{|c|}{$\sigma_{DT}$ =  0.0005655(7)  [nb] }\\
\hline
$|\Delta| >$ & $\sigma_{\Delta_{DT}}$ [nb]& $\sigma_{\Delta_{QH}}$ [nb]& $\sigma_{\Delta_{DT}}/\sigma_{DT}$& N$_{event}$\\
\hline
0.0001 & 0&0&0&0\\
\hline
0.00001& $3(1) \cdotp 10^{-10}$& $3(1) \cdotp 10^{-10}$&$5(2) \cdotp 10^{-7}$&6\\
\hline
0.000001&$1.2(2) \cdotp 10^{-9}$&$1.2(2) \cdotp 10^{-9}$&$2.1(4) \cdotp 10^{-6}$&26\\
\hline
\end{tabular}
\end{center}
\caption{Comparison between the codes based on the DT-method and
  QH-method for BaBar event selection cuts~(see Appendix~\ref{appex:KLOBA}). 
The contribution $\sigma_{\Delta}$ to the cross section $\sigma$  for a 
chosen $|\Delta|$. Subscripts $QH$ and $DT$ denote
 QT-method and DT-method respectively. $q^2 \in (0.34,0.96)$ GeV$^2$. See text
  for details on the definition of the Table entries. \label{test4}}
\end{table}

\begin{figure}[ht]%
      \centering
    \includegraphics[scale=0.6,angle=0]{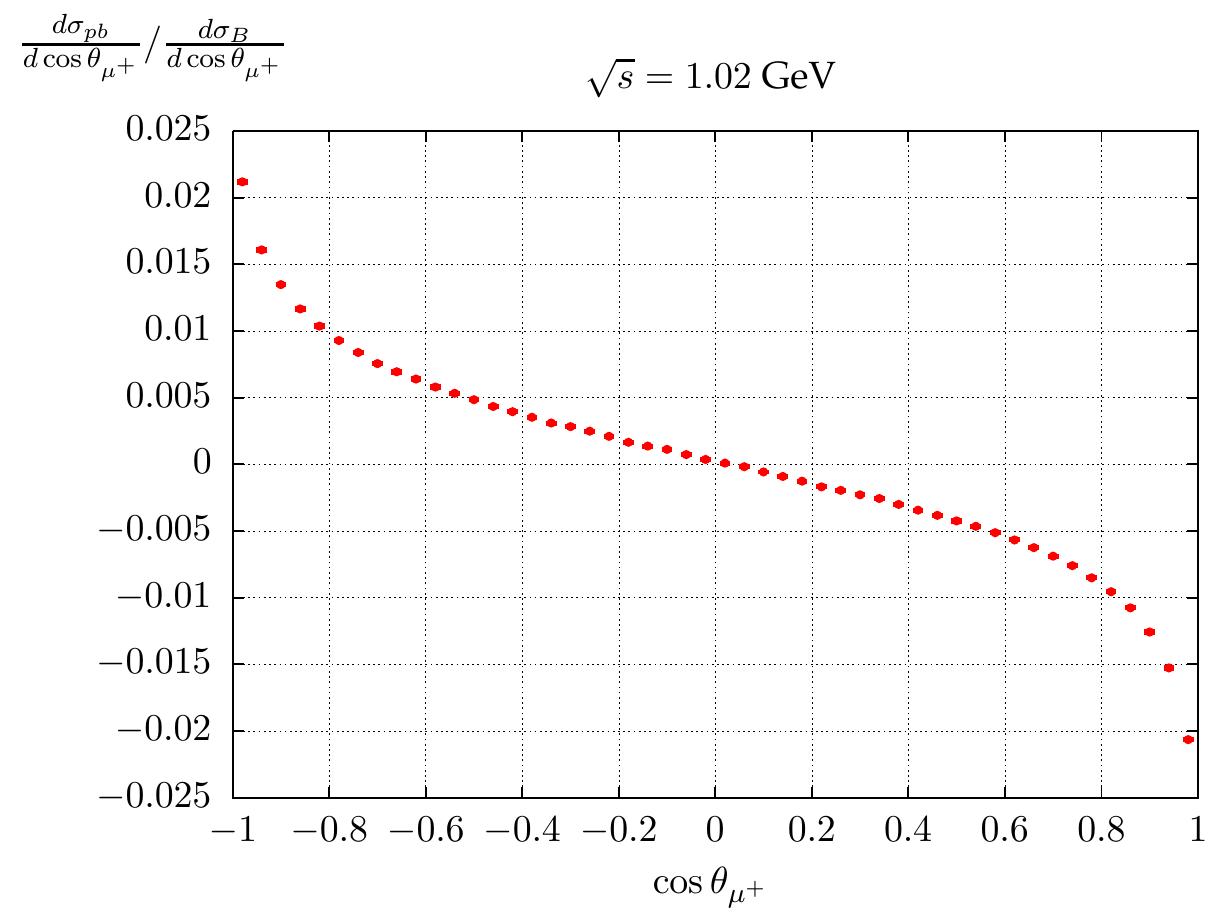} %
  \includegraphics[scale=0.6,angle=0]{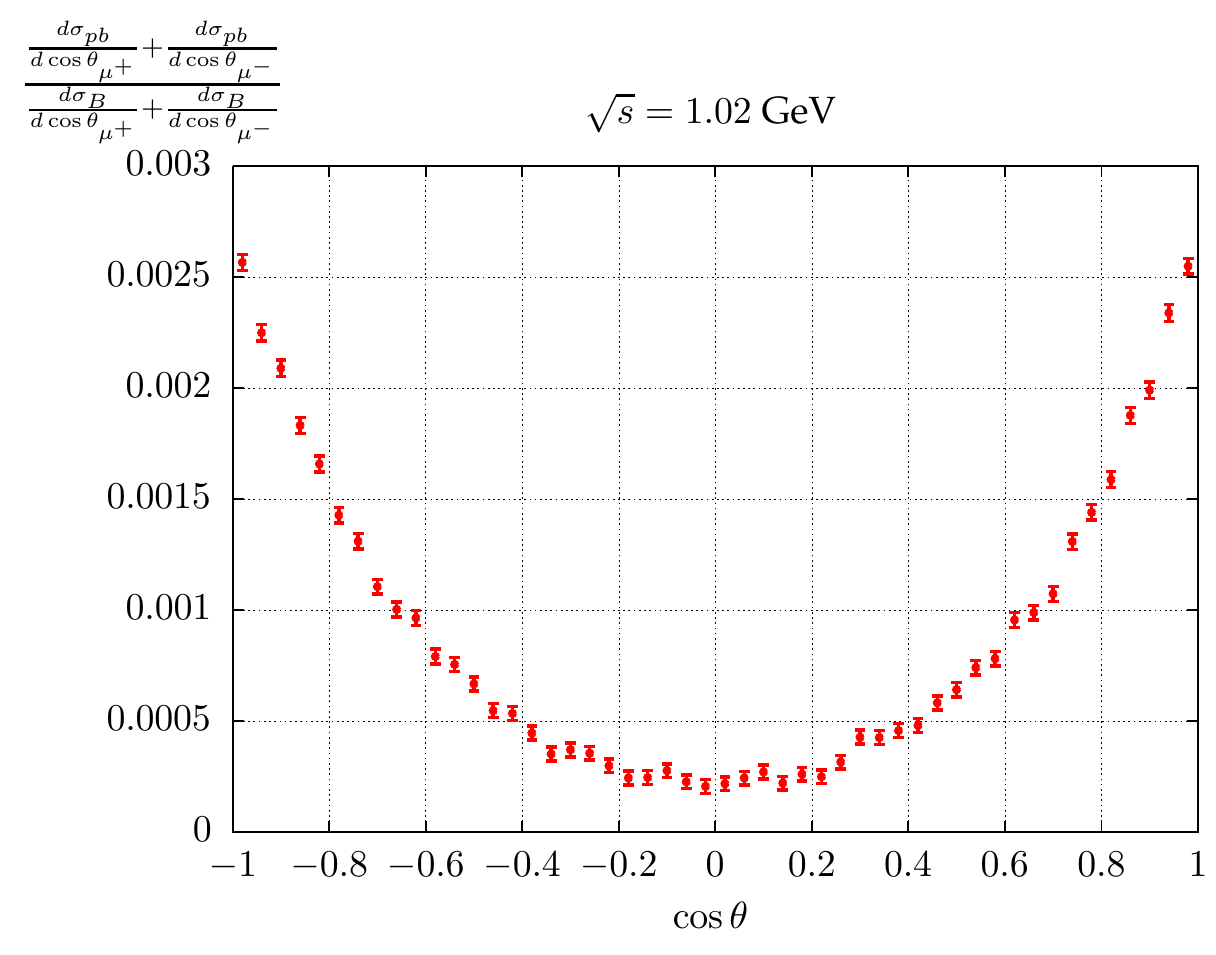} %
\\
   \includegraphics[scale=0.6,angle=0]{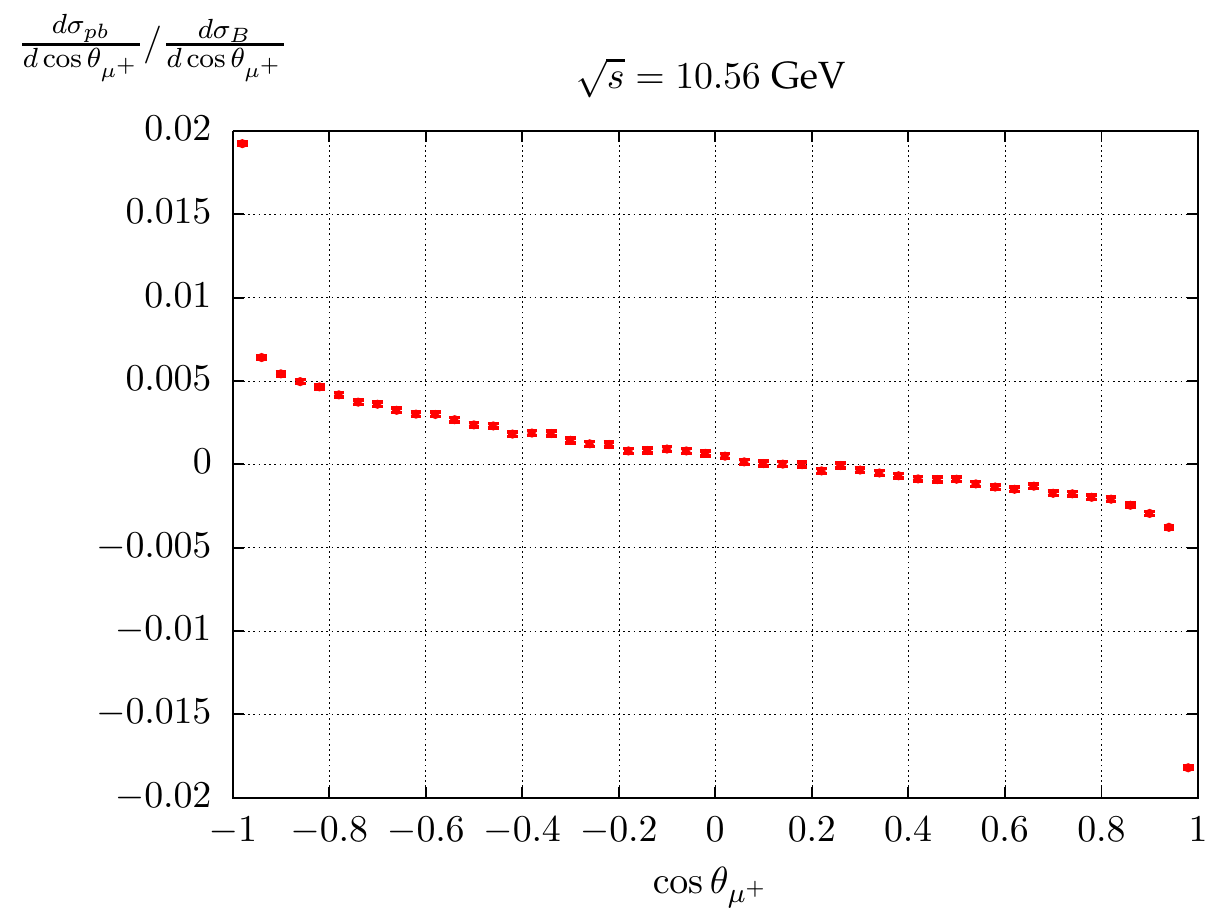} %
  \includegraphics[scale=0.6,angle=0]{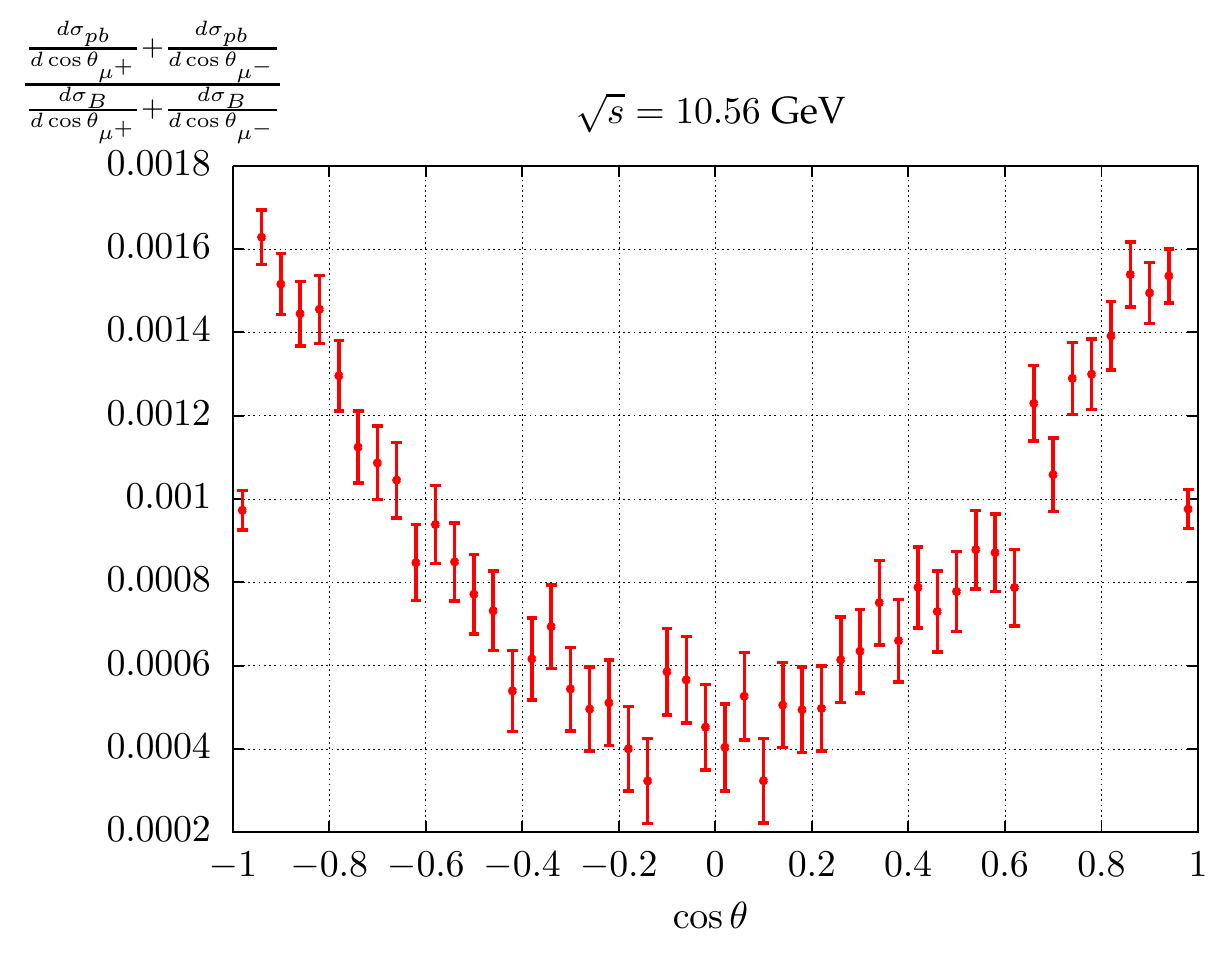} %
\caption{Relevance of NLO Penta-Box contributions at KLOE  (above) and Babar (below) energies for muon angular
distributions: $\theta_{\mu^+}$ is the $\mu^+$ polar angle, while $\theta$ is 
 the angle of $\mu^+$ or $\mu^-$ for the charge 'blind' observable. 
These definitions are the same in all the figures and will not be repeated
 in captions.   }
\label{fig:kloebabarcos}
  \end{figure}

\begin{figure}[ht]%
      \centering
       \includegraphics[scale=0.6,angle=0]{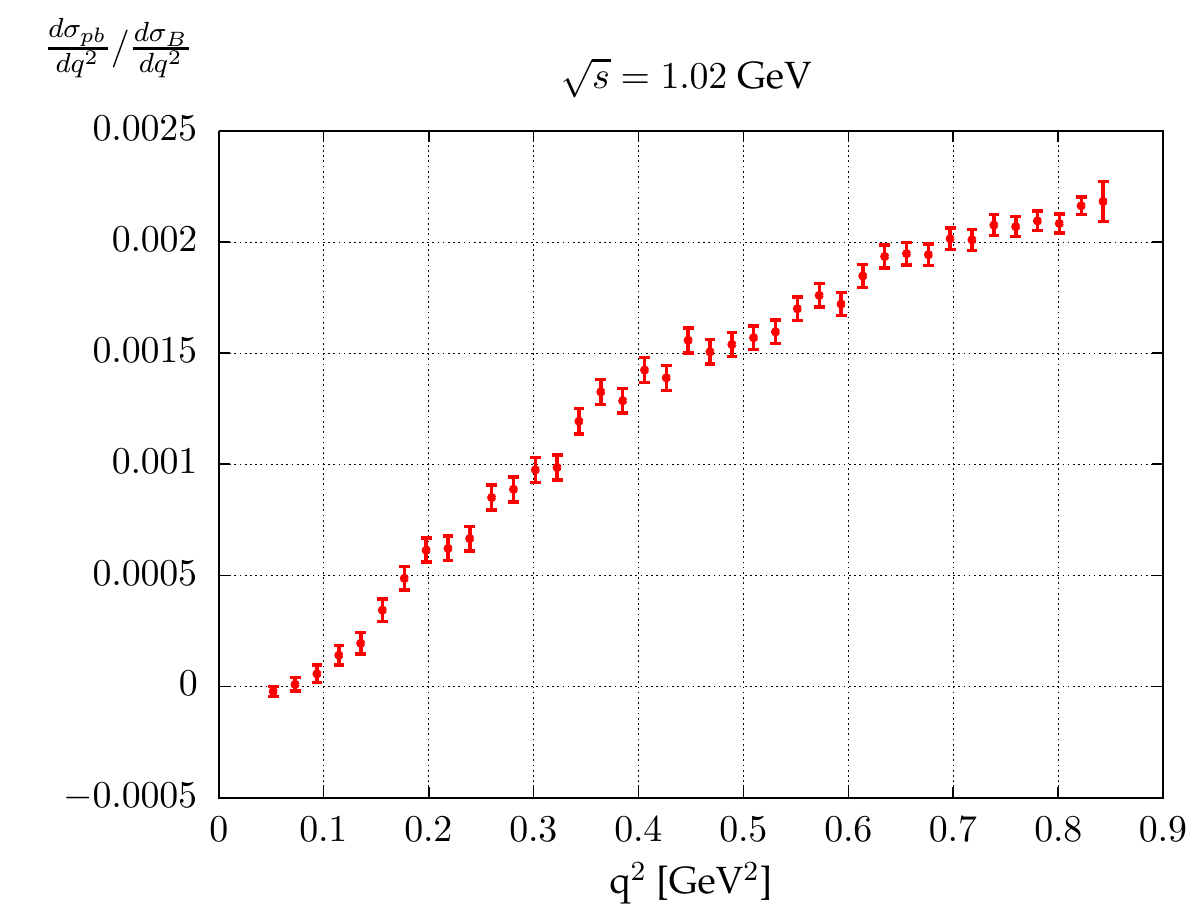} %
	  \includegraphics[scale=0.6,angle=0]{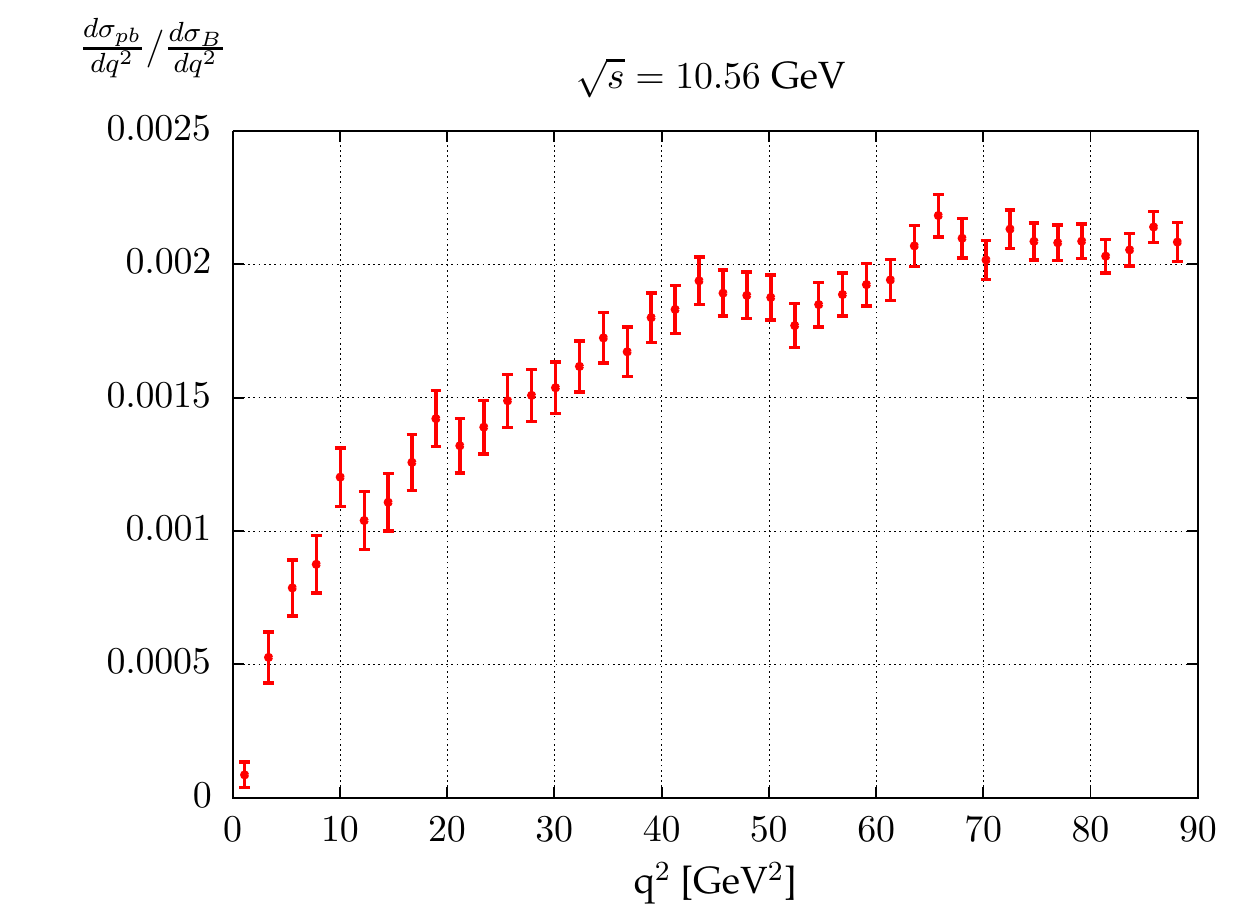} %
\caption{Relevance of NLO Penta-Box contributions at KLOE  (left) and Babar (right) energies for $q^2$ $\mu^+
\mu^-$ distributions.}
\label{fig:kloebabarqq}
  \end{figure}

The checks clearly show the control on the numerical accuracy of the
virtual corrections.  Additionally, using the DT-method  and realistic cuts (see Appendix~\ref{appex:KLOBA} for their definition) for KLOE and
Babar energies, we have studied the relevance of the most challenging
contribution in this article. For this purpose, we compare the contributions from one-loop Penta-Box diagrams
defined in Section~\ref{sec:class} with the Born contributions in
Fig.~\ref{fig:kloebabarcos} for muon angular distributions and in Fig.~\ref{fig:kloebabarqq} for
the $\mu^+ \mu^-$ invariant mass distribution. As we can clearly see from the
muon and antimuon angular distributions, the size of the Penta-Box contributions
can reach the percent level and they cannot be neglected for the charge
 odd observables. We confirm here the expectations that
 the neglected corrections~\cite{Czyz:2004rj} for the charge even distributions
 are indeed small.
 For a classification of the charge odd and even contributions, we refer the reader
 to Ref.~\cite{Czyz:2003ue}, where it was done for charged pions in the final state.
 Replacing pions with muons does not change the classification presented there.


  
\begin{figure}[ht]%
  \hskip 0.5cm  \vskip 0.55cm \includegraphics[scale=0.35,angle=0]{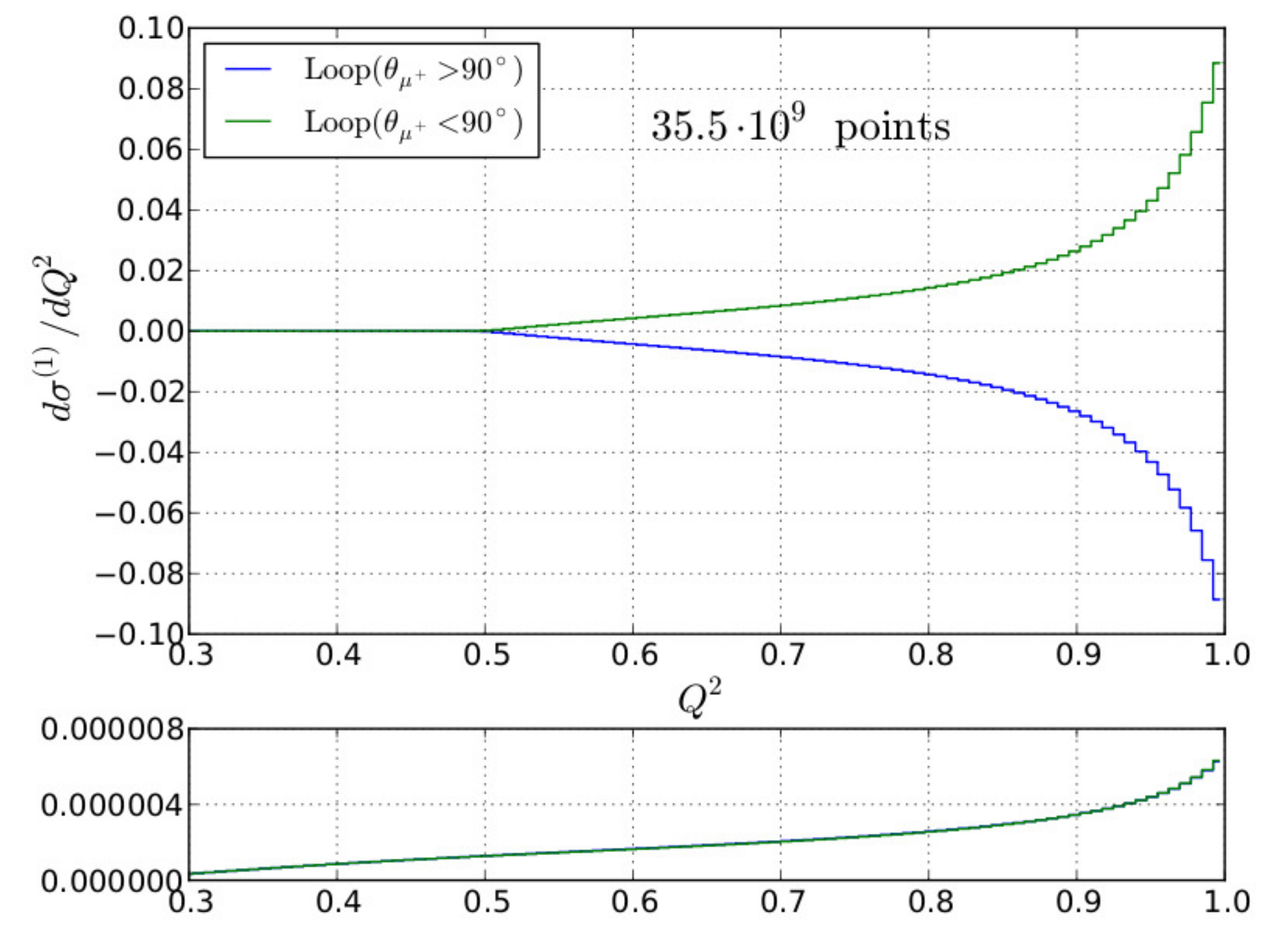} \vskip -5.75cm  \hskip 7.5 cm
\includegraphics[scale=0.31,angle=270]{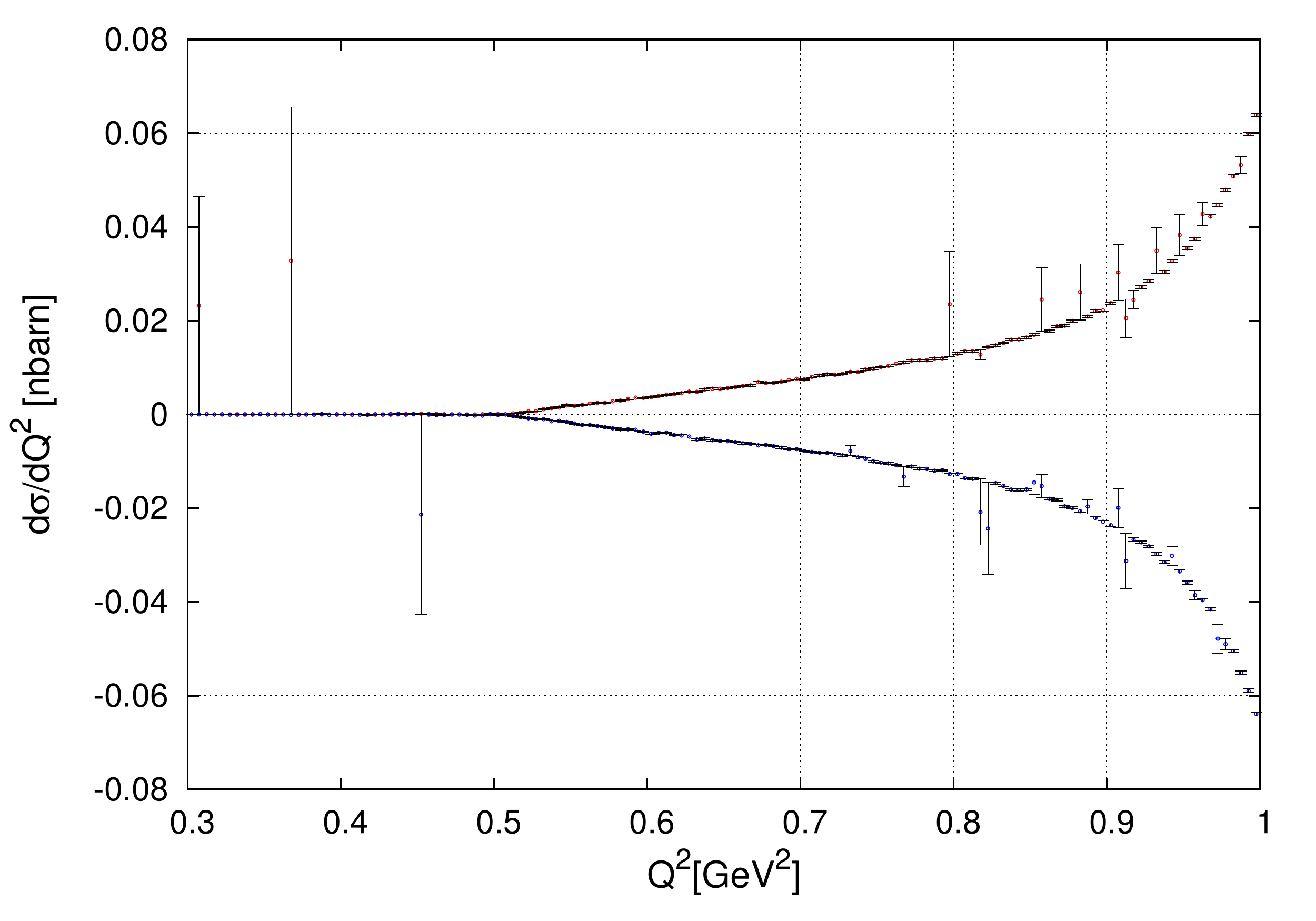}
\caption{On left: Muon pair distributions including 5-point functions at KLOE
  calculated with \texttt{PJFry} (bottom: absolute error estimate). On right:
  the same calculated without decicated routines to avoid small Gram determinants.
Approximately $4\cdot 10^{10}$ ($10^9$) events have been generated.
 }
\label{fig:acc}
  \end{figure}

 To appreciate these results,   
in Fig.~\ref{fig:acc}, we plot the Penta-Box contributions using the
\texttt{PJFry} package with (left) and without (right) using expansion for small Gram
determinants. The right panel reveals discrepancies only after
increasing the number of Monte Carlo events to $10^9$~\cite{kajphd,Gluza:2012yz}.
\texttt{PJFry} treats properly 
 small Gram determinants, as discussed in details in~\cite{yundin-phd-2012--oai:export}.  
  With the \texttt{PJFry} package, the leading  inverse Gram determinants
  $|G^{(5)}|$ are eliminated in the tensor reduction and 
small inverse Gram determinants $|G^{(4)}|$ are avoided using asymptotic expansions and Pade approximants. 
For more details concerning the numerical stability of the tensor reductions, see
Refs.~\cite{yundin-phd-2012--oai:export,Fleischer:2011bi}. The results are
completely stable and well under control. 

 
The soft real emission analytic formulae were also checked. Firstly, by
comparing to the integral obtained by means of a Monte Carlo methods. A good agreement was found even if the
method is limited in some cases to an
accuracy of  $2\cdot 10^{-4}$. Secondly, the numerical stability of the code
 was tested comparing the quadruple and the double precision versions of the same
 code. The relative accuracy of the double precision version used 
 in the released code
 of the generator is at the level of $10^{-7}$ at 1~GeV, while at 10~GeV , it
 was only about $10^{-3}$ in some corners of the phase space. However, since
 those phase space regions did not
 affect the relevant observables (invariant mass and angular 
 distributions), the code was not changed to cure this behaviour by means of appropriate expansions. 

 The new contributions of the real two-photon emission
 were tested comparing two completely independent codes.
  In one, the trace method (T) and FORM~\cite{Kuipers:2012rf}
  was used to obtain an analytic result,
  in the second one, the helicity amplitude method (H),
  described in~\cite{Rodrigo:2001kf} and in Appendix~\ref{appex:twophoton},
 was applied.

   The biggest observed relative difference of the codes was at the level of $10^{-4}$
  even if both codes were using double precision only. Additionally, in both
  cases gauge invariance was checked. For the T-method
  analytically, and for the H-method numerically, obtaining a relative accuracy of $10^{-15}$.

 Both the soft and the real parts were tested checking the independence 
  of the cross section and differential distributions of the separation parameter between the soft part,
where the integral
  over the one photon phase space is performed analytically, and the hard
 part, where the integral is obtained using the Monte Carlo generation.  The accuracy of this test was
 $2\cdot 10^{-4}$. A perfect agreement at this level of accuracy 
 was observed.

\section{Impact of the radiative corrections added
to the event generator on the  pion form factor measurements at BaBar and KLOE\label{sec-impact}}

\begin{figure}[h!]
\begin{center}
\vspace*{-4.0cm}
\hspace*{-2.0cm}\subfloat{\includegraphics[width=.8\textwidth]{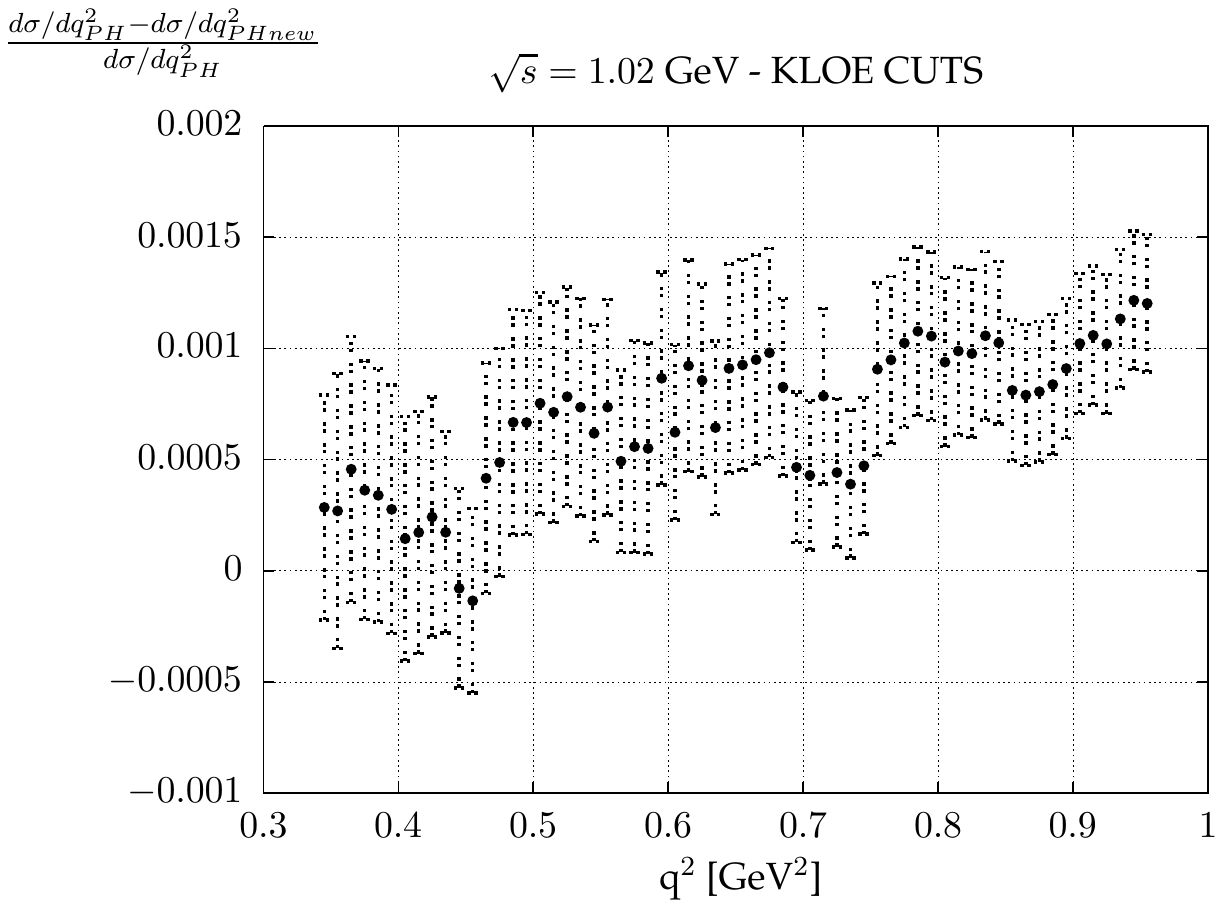}}
\hspace*{-4.0cm}\subfloat{\includegraphics[width=.8\textwidth]{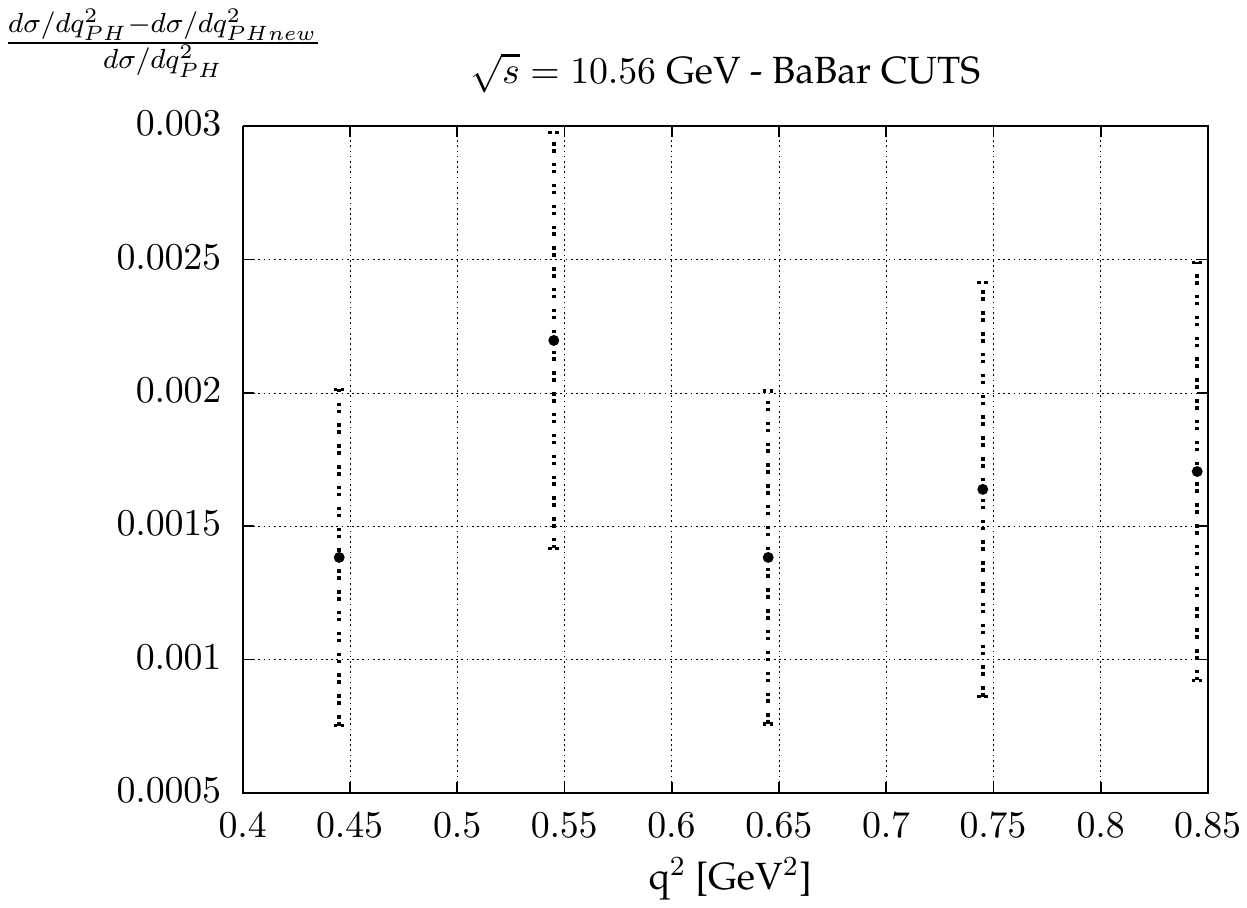}}
\end{center}
\vspace*{-9.0cm}
\caption[]{The relative difference between differential cross sections of PHOKHARA7.0 
 and PHOKHARA9.0 (with subscript `new').
 }
\label{q2reldif}
\end{figure}

 PHOKHARA7.0 has been used by BaBar and KLOE until quite recently. In fact, from version 4.0 to 7.0 the
muon production channel was not changed. 
Comparing numerics with PHOKHARA9.0, which includes  the
 complete NLO corrections, one has to distinguish between the charge average distributions 
 for which the bulk of the NLO corrections was already included in PHOKHARA7.0 and the charge sensitive observables 
for which version 7.0 was limited to the leading order only. In the experimental framework for the extraction
 of the hadronic cross section,  charge averaged observables were used. 
The most important invariant muon pair mass distribution (i.e. the  $q^2$ distribution), from
which the hadronic cross section is extracted, is shown in Fig.~\ref{q2reldif}.
 As one can see, the radiative corrections missing in  PHOKHARA7.0 are small. They reach up to 0.1~\%
 for the KLOE event selection and up to 0.25\% for the BaBar event selection. 

\begin{figure}[t!]
\begin{center}
\includegraphics[width=.49\textwidth]{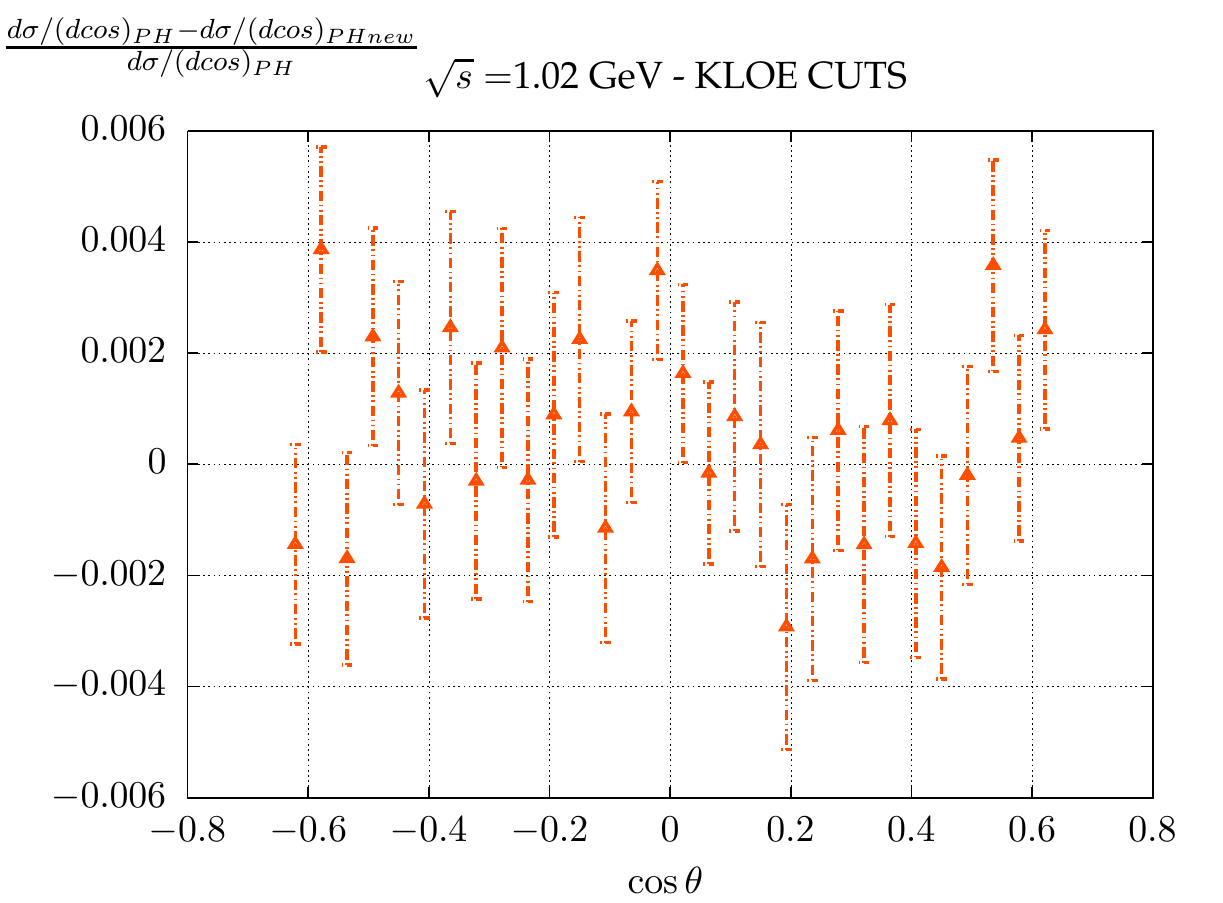}
\includegraphics[width=.49\textwidth]{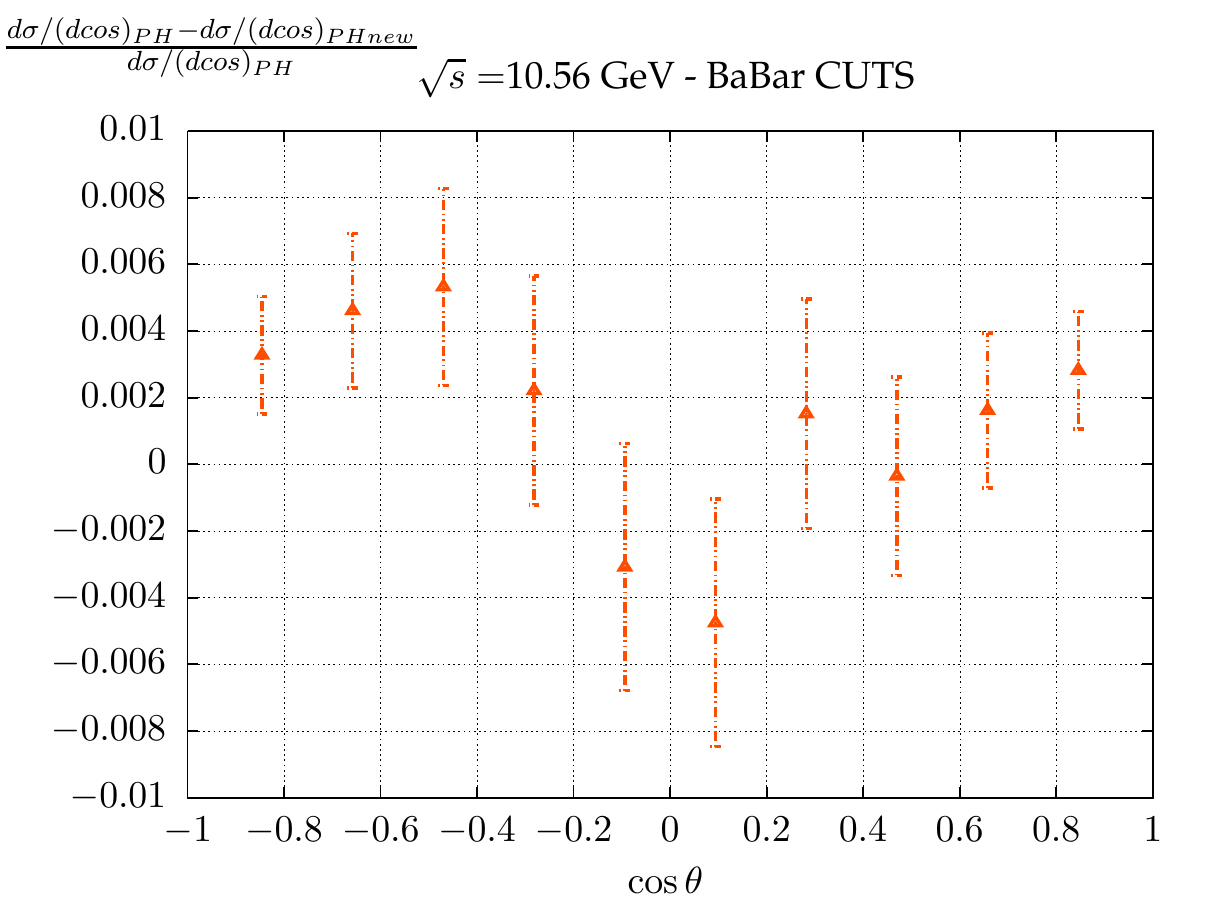}
\end{center}
\caption[]{The relative difference between differential cross sections of PHOKHARA7.0 
 and PHOKHARA9.0 (with subscript `new'). $q^2\in (0.54,0.55)$.
 }
\label{ang1reldif}
\end{figure}

\begin{figure}[ht]
\begin{center}
\includegraphics[width=.49\textwidth]{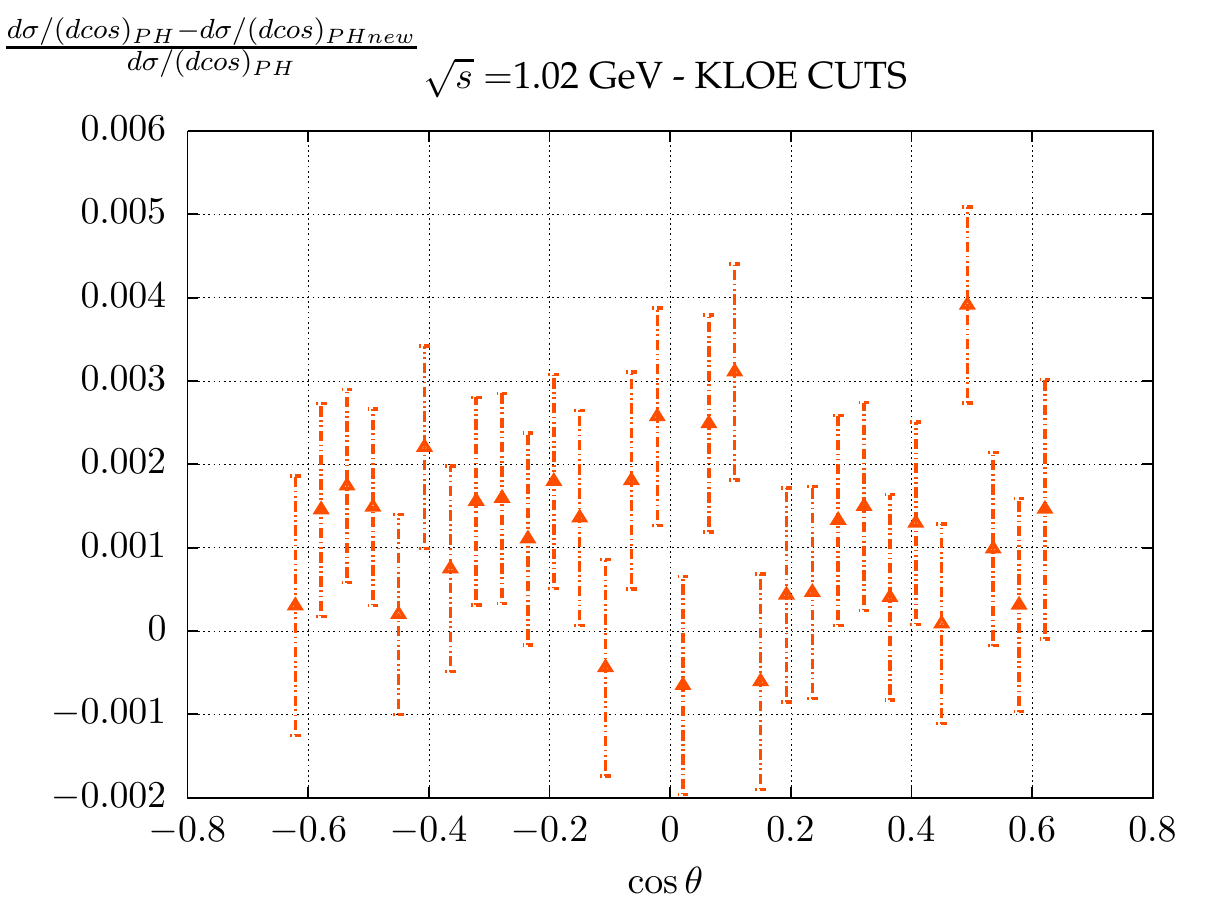}
\includegraphics[width=.49\textwidth]{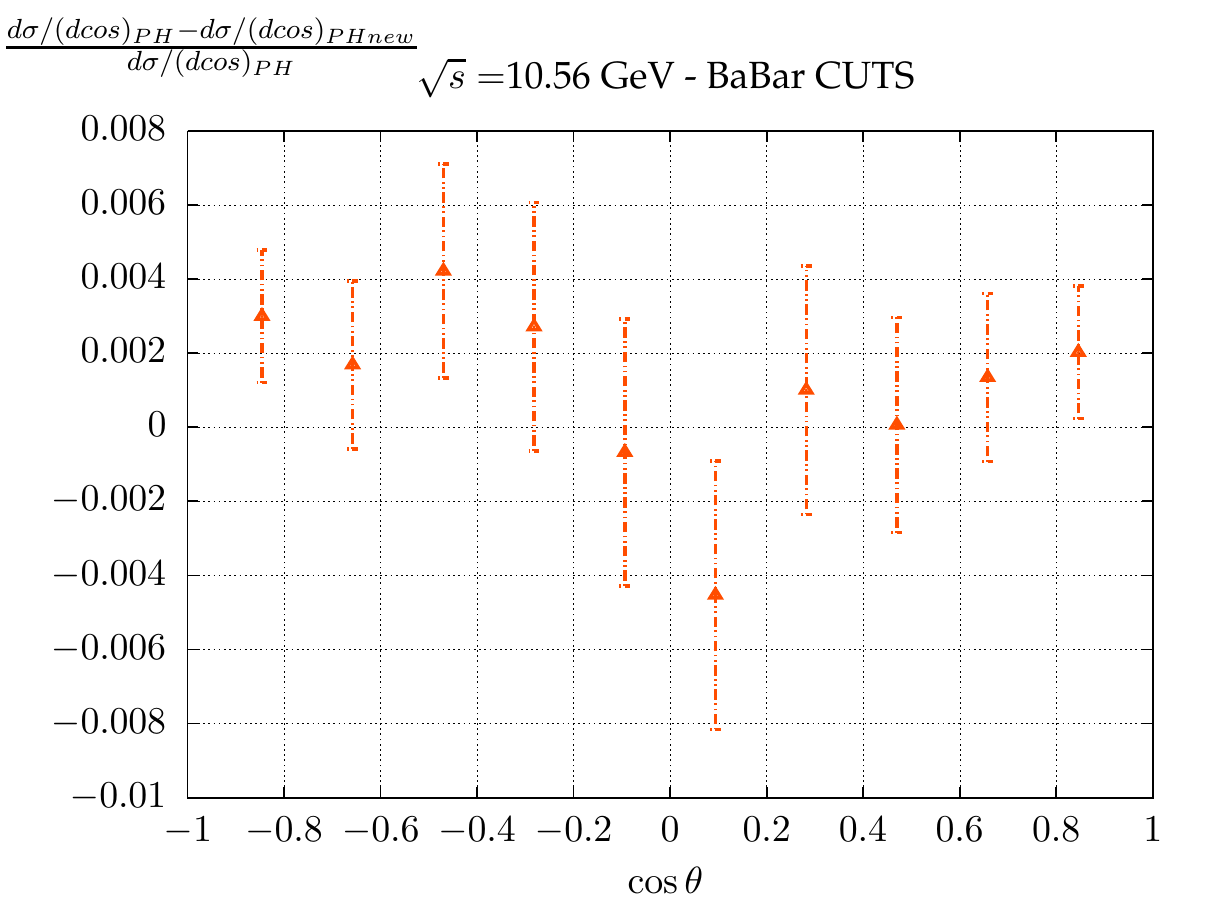}
\end{center}
\caption[]{The relative difference between differential cross sections of PHOKHARA7.0 
 and PHOKHARA9.0 (with subscript `new'). $q^2\in (0.94,0.95)$ for KLOE cuts;
$q^2\in (0.74,0.75)$ for BaBar cuts.
 }
\label{ang1reldif1}
\end{figure}

The charge averaged angular 
 distributions are also not very much different as shown in Fig.~\ref{ang1reldif} and Fig.~\ref{ang1reldif1}
 for different $q^2$ bins. For other muon pair invariant mass ranges, the results are similar to the ones
 shown. We can conclude at this point that in the experiments using the charge averaged observables, 
 the missing radiative corrections are very small and should not have affected the extraction
 of the hadronic cross section.
  
\begin{figure}[th!]
\begin{center}
\includegraphics[width=.49\textwidth]{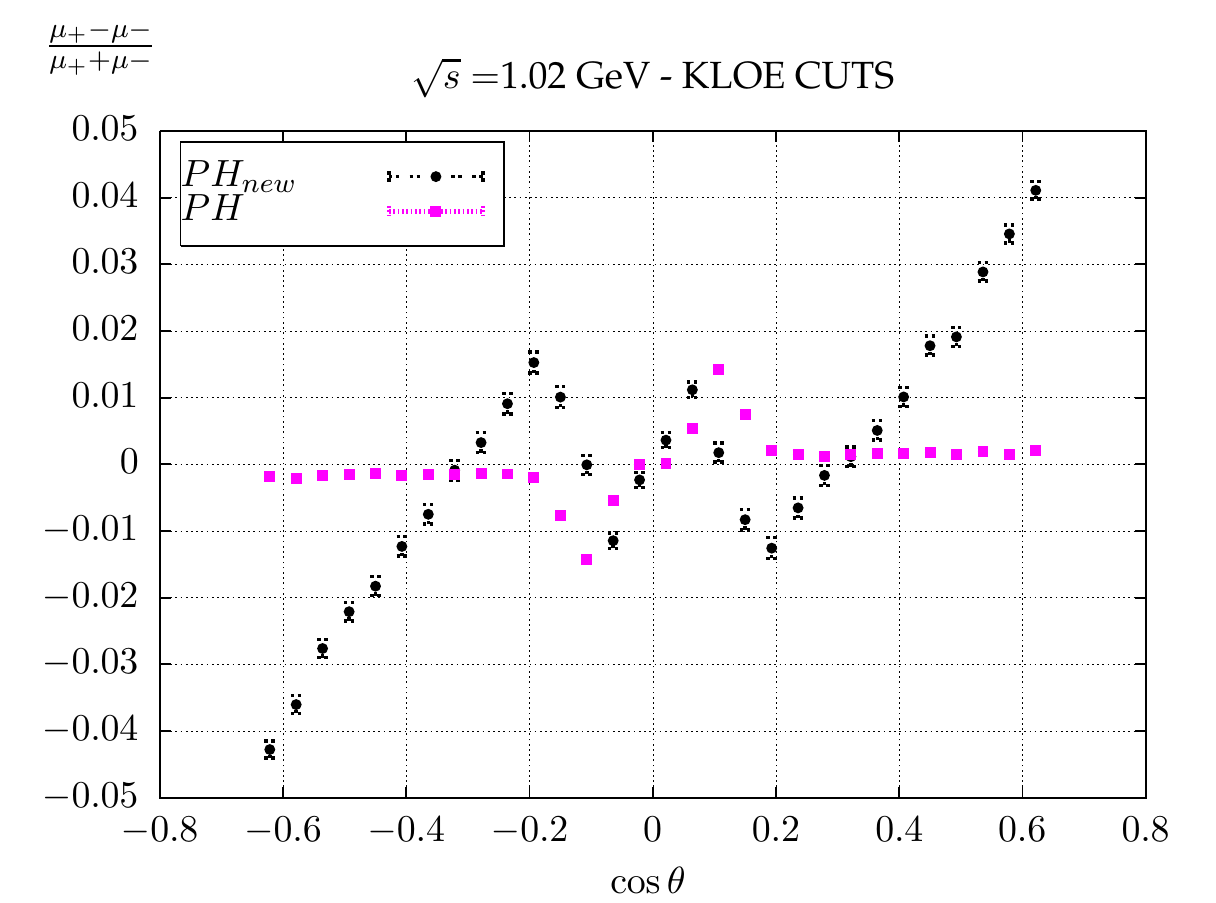}
\includegraphics[width=.49\textwidth]{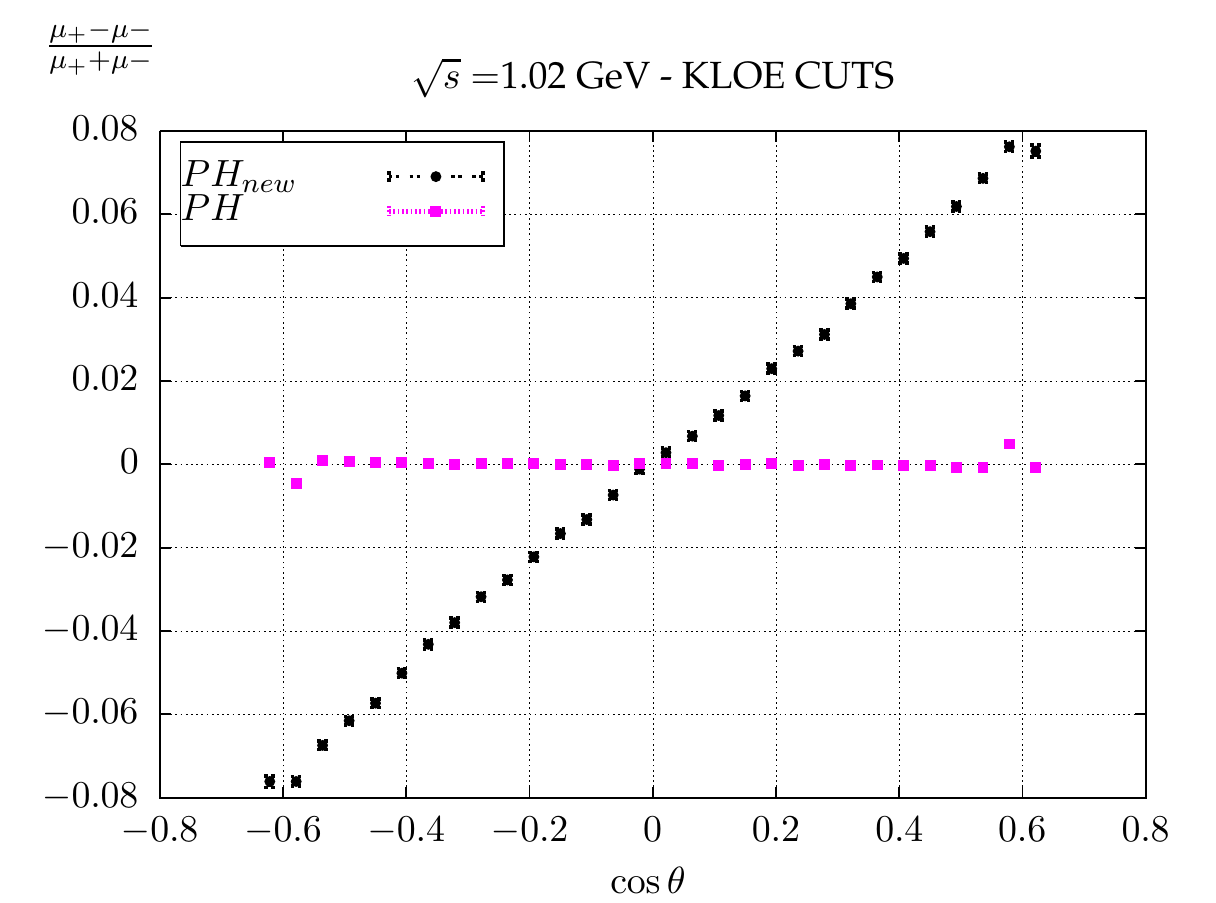}
\end{center}
\caption[]{The asymmetries given by
  PHOKHARA7.0 (denoted as $PH$)
 and PHOKHARA9.0 (denoted as $PH_{new}$). $q^2\in (0.54,0.55)$ - left plot;
 $q^2\in (0.94,0.95)$ - right plot.
 }
\label{asym}
\end{figure}

\begin{figure}[th!]
\begin{center}
\includegraphics[width=.49\textwidth]{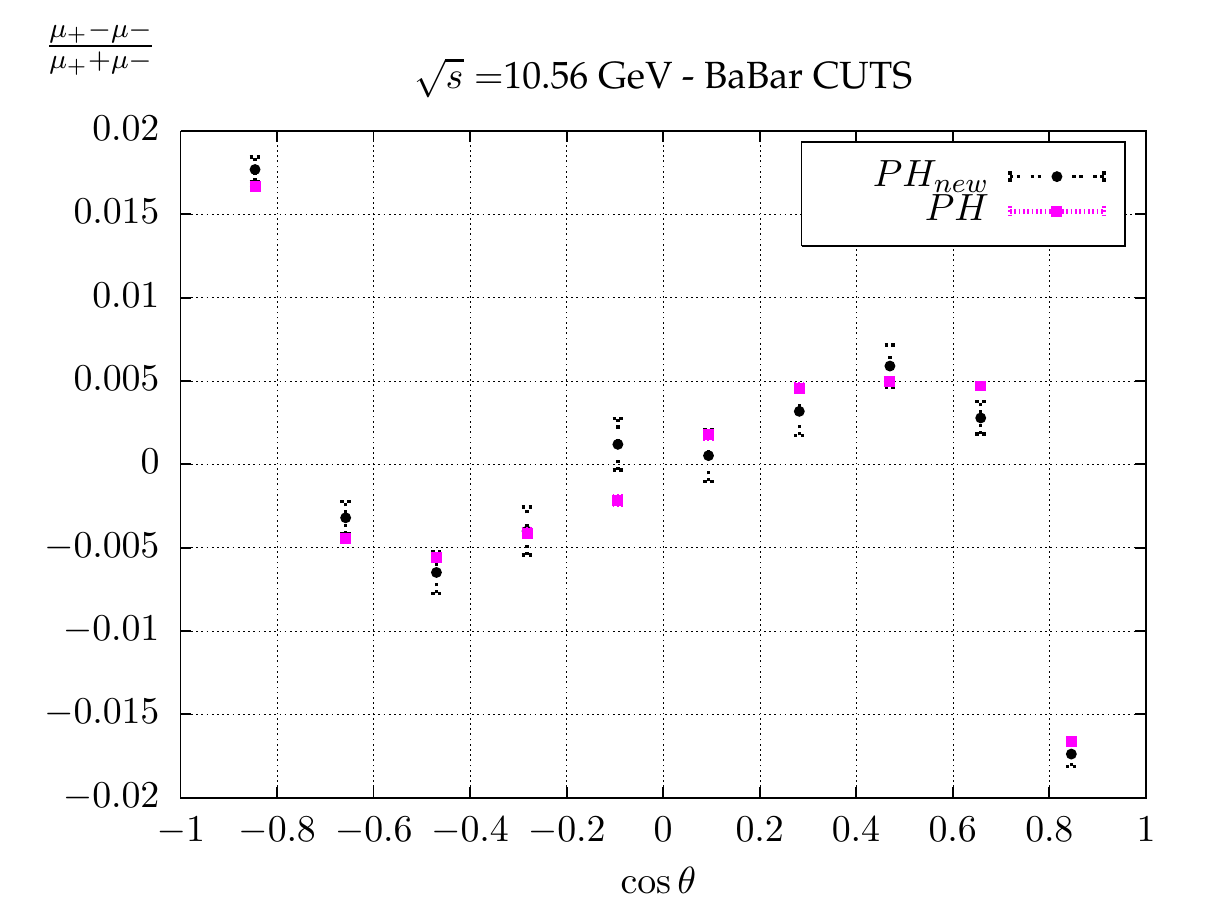}
\includegraphics[width=.49\textwidth]{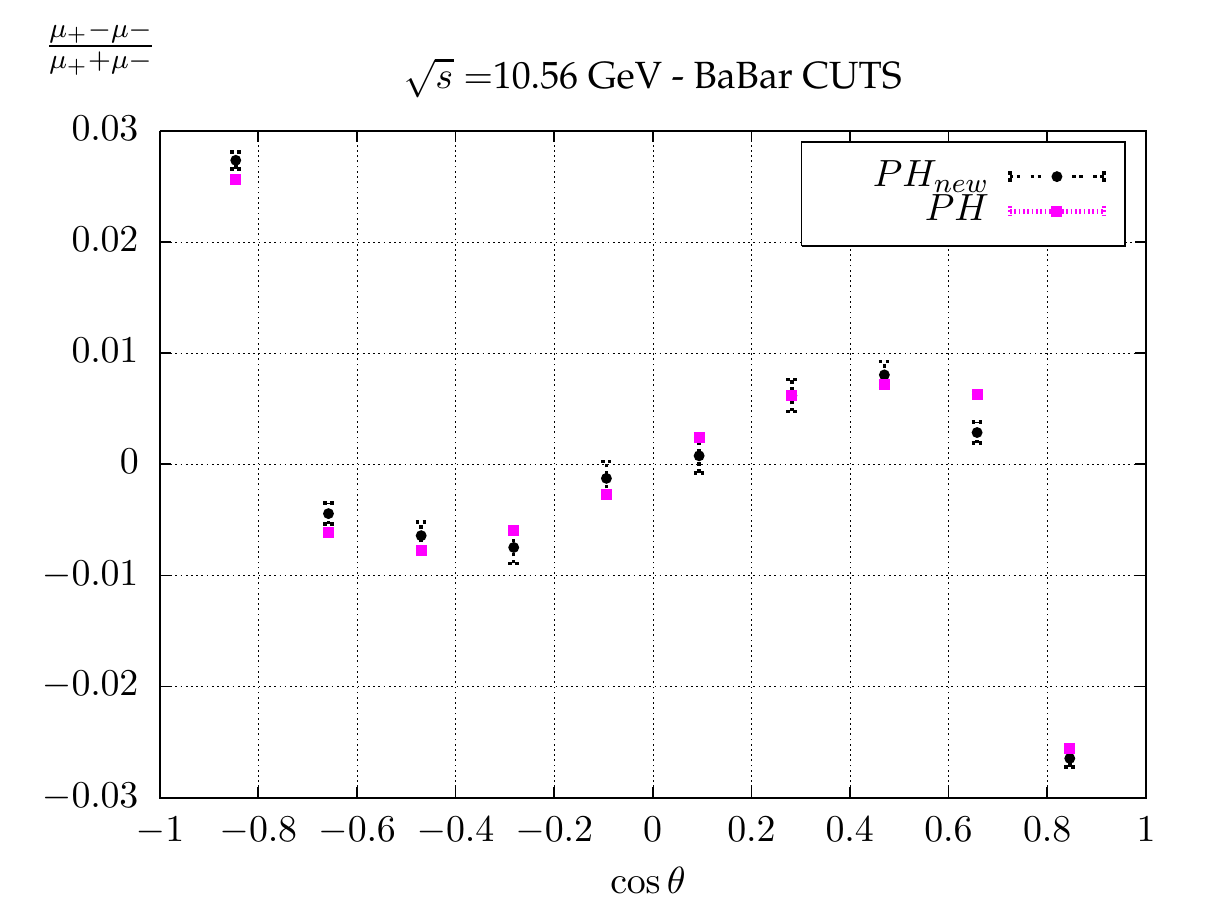}
\end{center}
\caption[]{The asymmetries given by
  PHOKHARA7.0 (denoted as $PH$)
 and PHOKHARA9.0 (denoted as $PH_{new}$). $q^2\in (0.54,0.55)$ - left plot;
 $q^2\in (0.74,0.75)$ - right plot.
 }
\label{asym1}
\end{figure}

For the charge sensitive observables, which were available at PHOKHARA7.0 only at LO 
 (the ISR-FSR interference was present at LO only) the new corrections are relatively bigger 
 and reach typically a few percent as expected from NLO corrections. The KLOE event selection
  was designed to diminish the FSR radiative corrections and as such it was also mostly killing 
  the asymmetry coming from one photon emission. The asymmetry coming from the two photon emission
 is however surviving the cuts as shown in Fig.~\ref{asym}. For BaBar, the asymmetry
 is naturally suppressed by tagging the photon at large angles. At low invariant masses, as
 compared to the energy available at the experiment, it forces the muons to fly in the opposite
 direction to the photon and thus the suppression. The asymmetry is at the level of  few percent and it
 is dominated by the LO contributions as shown in Fig.~\ref{asym1}. 

\section{Conclusions  \label{sec-conc}}
     The presented studies allow for the development of a numerically
  stable Monte Carlo event generator PHOKHARA9.0 simulating the reaction
  $e^+e^- \to \mu^+\mu^-\gamma$ with full NLO QED accuracy. 
The radiative corrections which were missing in the previous
  versions of the generator can reach a few percent.
Though, it was shown that the charge blind observables used
 by the BaBar and KLOE collaborations are affected only at the level of 
 0.1\% for KLOE and 0.3\% for BaBar. 
We conclude that the observed discrepancies
  between these experiments cannot be attributed to the missing corrections for  the reaction  $e^+e^- \to
\mu^+\mu^-\gamma$ in PHOKHARA4.0~\cite{Czyz:2004rj,Czyz:2004ua} to PHOKHARA8.0~\cite{Czyz:2013xga} .

 \acknowledgments{This work has been supported by the
Research Executive Agency (REA) of the European Union under
the Grant Agreement number PITN-GA-2010-264564 (LHCPhenoNet),
by the Polish Ministry of Science and High Education
under grant number N N202 102638,
by the Spanish Government and EU ERDF funds
(grants FPA2011-23778, FPA2011-23596 and CSD2007-00042
Consolider Project CPAN), by GV (PROMETEUII/2013/007) and by the Deutsche Forschungsgemeinschaft via the
Sonderforschungsbereich/Transregio SFB/TR-9 Computational Particle Physics. This work is  part of the activity of the ``Working Group on Radiative
 Corrections and Monte Carlo Generators for Low Energies"  
 [\url{http://www.lnf.infn.it/wg/sighad/}].
 FC is funded by a Marie Curie fellowship (PIEF-GA-2011-298960).
MG was supported by ``\'Swider'' PhD program.
}


\appendix
\section{KLOE and BaBar event selections cuts
   \label{appex:KLOBA}}

\subsection{KLOE}
\begin{itemize}
 \item $\sqrt{s} = 1.02$~GeV
 \item Muon tracks: $50^{\circ} < \theta_{\mu^\pm} < 130^{\circ}$
 \item Missing photon angle $<15^{\circ} (>165^{\circ})$
 \item Track mass: $80$~MeV~$< m_{\rm trk} < 115$~MeV 
 \item $q^2 \in (0.34, 0.96)$
\end{itemize}

\subsection{BaBar}
\begin{itemize}
 \item $\sqrt{s} = 10.56$~ GeV
 \item Muon tracks: $20^{\circ} < \theta_{\mu^\pm} < 160^{\circ}$
 \item Minimal photon energy/missing energy 3 GeV
 \item $|q_1|>1$~GeV (antimuon) and $|q_2|>1$~GeV (muon)
 \item $q^2 \in (0.34, 0.96)$
\end{itemize}

\section{Two-hard photon emission  \label{appex:twophoton}}
We use here the following notation:
\begin{itemize}
 \item $p_1$ - positron ($e^+$) four momenta;
 \item $p_2$ - electron ($e^-$) four momenta;
 \item $q_1$ - antimuon ($\mu ^+$) four momenta;
 \item $q_2$ - muon ($\mu ^-$) four momenta;
 \item $k_1$, $k_2$ - photon four momenta.
\end{itemize}
The coefficients of the two-hard photon emission amplitude of Eq.~\ref{eq:a1}
are given by
\begin{eqnarray}
A\left(\lambda_1,\lambda_2\right) = \frac{-e^{3}}{4s} ( \frac{a_1 \varepsilon ^{*-}_{2} a_3}{(k_{1}\cdotp q_{1}) N_1} +\frac{a_1 \varepsilon ^{*-}_{1}a_5}{(k_{2}\cdotp q_{1}) N_1} + \frac{a_7 \varepsilon ^{*-}_{1} a_9}{(k_{2}\cdotp q_{2}) N_2}\\ \nonumber
+ \frac{a_{11}\varepsilon ^{*-}_{2} a_9}{(k_{1}\cdotp q_{2}) N_2} - \frac{a_7 J_{e^{+}e^{-}}^{-} a_3 }{(k_{1}\cdotp q_{1})(k_{2}\cdotp q_{2})} - \frac{a_{11} J_{e^{+}e^{-}}^{-} a_5 }{(k_{2}\cdotp q_{1})(k_{1}\cdotp q_{2})})
\end{eqnarray}
and
\begin{eqnarray}
B\left(\lambda_1,\lambda_2\right) = \frac{-e^{3}}{4s} ( \frac{a_2 \varepsilon ^{*+}_{2} a_4}{(k_{1}\cdotp q_{1}) N_1} + \frac{a_2 \varepsilon ^{*+}_{1}a_6}{(k_{2}\cdotp q_{1}) N_1}+ \frac{a_8 \varepsilon ^{*+}_{1} a_{10}}{(k_{2}\cdotp q_{2}) N_2}\\ \nonumber
+ \frac{a_{12}\varepsilon ^{*+}_{2} a_{10}}{(k_{1}\cdotp q_{2}) N_2}- \frac{a_8 J_{e^{+}e^{-}}^{+} a_4
}{(_{1}\cdotp q_{1})(k_{2}\cdotp q_{2})}- \frac{a_{12} J_{e^{+}e^{-}}^{+} a_6 }{(k_{2}\cdotp
q_{1})(k_{1k}\cdotp q_{2})}),
\end{eqnarray}
where
\begin{alignat}{2}
\label{eqlong}
N_1 &= k_{1}\cdotp q_{1} +  k_{2}\cdotp q_{1} +  k_{1}\cdotp k_{2} , & ~~~~~~
N_2 &= k_{1} \cdotp q_{2} +  k_{2}\cdotp q_{2} +  k_{1}\cdotp k_{2},
\\ \nonumber
a_1 &= J_{e^{+}e^{-}}^{-}(p_1^+ + p_2^+)-2q_{2}\cdotp J_{ e^{+}e^{-}} , & ~~~~~~
a_2 &= J_{e^{+}e^{-}}^{+}(p_1^- + p_2^-)-2q_{2}\cdotp J_{ e^{+}e^{-}},
\\ \nonumber
a_3 &= k^{+}_{1}\varepsilon ^{*-}_{1}+2(\varepsilon ^{*}_{1} \cdotp q_{1}) , &
a_4 &= k^{-}_{1}\varepsilon ^{*+}_{1}+2(\varepsilon ^{*}_{1} \cdotp q_{1}) ,
\\ \nonumber
a_5 &= k^{+}_{2}\varepsilon ^{*-}_{2}+2(\varepsilon ^{*}_{2} \cdotp q_{1}) , &
 a_6 &= k^{-}_{2}\varepsilon ^{*+}_{2}+2(\varepsilon ^{*}_{2} \cdotp q_{1}) , 
\\ \nonumber
a_7 &= \varepsilon ^{*-}_{2} k^{+}_{2} +2(\varepsilon ^{*}_{2} \cdotp q_{2}) , &
 a_8 &= \varepsilon ^{*+}_{2} k^{-}_{2} +2(\varepsilon ^{*}_{2} \cdotp q_{2}),
\\ \nonumber
a_9 &= (p_1^+ + p_2^+)J_{e^{+}e^{-}}^{-} -2q_{1}\cdotp J_{ e^{+}e^{-}} , &
  a_{10}& = (p_1^- + p_2^-)J_{e^{+}e^{-}}^{+} -2q_{1}\cdotp J_{ e^{+}e^{-}},
\\ \nonumber
a_{11} &= \varepsilon ^{*-}_{1} k^{+}_{1} +2(\varepsilon ^{*}_{1} \cdotp q_{2}) , &
a_{12} &=\varepsilon ^{*+}_{1} k^{-}_{1} +2(\varepsilon ^{*}_{1} \cdotp q_{2}) .  
\\ \nonumber
\end{alignat}
For the reader's convenience, we give
 here all the relevant definitions. The gamma matrices and related objects are defined in the following form:

\begin{align}
\gamma^{\mu}= \left( 
\begin{array}{cc}
0 & \sigma^{\mu}_+ \\
\sigma^{\mu}_- & 0
\end{array} \right)~, \qquad \mu=0,1,2,3~, \qquad 
\ta{a} = a_{\mu} \gamma^{\mu} = \left( 
\begin{array}{cc}
0   & a^+ \\
a^- & 0
\end{array} \right)~,
\end{align}

\begin{align}
a^{\pm} =  a^{\mu} \sigma_{\mu}^{\pm} &= \left( \begin{array}{cc}
a^0 \mp a^3      & \mp (a^1 - i a^2) \\
\mp (a^1 + i a^2) & a^0 \pm a^3
\end{array} \right)~.
\end{align}

The helicity spinors $u$ and $v$ for a particle and an antiparticle 
are given by:
\begin{align}
u(p,\lambda) &= \left( \begin{array}{c}
\sqrt{E-\lambda |{\bf p}|}\; \chi({\bf p},\lambda) \\
\sqrt{E+\lambda |{\bf p}|}\; \chi({\bf p},\lambda)
\end{array} \right) \equiv \left( \begin{array}{c}
u_I \\ u_{II} \end{array} \right)~, \non \\ 
v(p,\lambda) &= \left( \begin{array}{c}
-\lambda \sqrt{E+\lambda |{\bf p}|}\; \chi({\bf p},-\lambda) \\
\lambda \sqrt{E-\lambda |{\bf p}|}\; \chi({\bf p},-\lambda)
\end{array} \right) \equiv \left( \begin{array}{c}
v_I \\ v_{II} \end{array} \right)~.
\end{align}Where helicity $\lambda/2=\pm 1/2$.

The helicity eigenstates $\chi({\bf p},\lambda)$ are expressed
in terms of the polar and azimuthal angles of the momentum
vector ${\bf p}$:
\begin{align}
\chi({\bf p},+1) &= \left( \begin{array}{r}
\cos{(\theta/2)} \\ e^{i\phi} \sin{(\theta/2)}
\end{array} \right)~, \non \\ 
\chi({\bf p},-1) &= \left( \begin{array}{r}
-e^{-i\phi}\sin{(\theta/2)} \\ \cos{(\theta/2)}
\end{array} \right)~.
\end{align}

However, for incoming particles in their CMS coordinate frame
 with z-axis along the positron direction they simplify to:
\begin{align}
\chi({\bf p_1},+1) = \left( \begin{array}{r}
1 \\ 0
\end{array} \right)~, \qquad
\chi({\bf p_1},-1) = \left( \begin{array}{r}
0 \\ 1
\end{array} \right)~,
\end{align}for positron and 

\begin{align}
\chi({\bf p_2},+1) = \left( \begin{array}{r}
0 \\ -1
\end{array} \right)~,  \qquad
\chi({\bf p_2},-1) = \left( \begin{array}{r}
1 \\ 0
\end{array} \right)~,
\end{align}for electron.

The photon polarisation vectors in the helicity basis are defined as 
\begin{align}
\varepsilon^{\mu}(k_i,\lambda_i=\mp) = \frac{1}{\sqrt{2}} \big(  0,
\pm \cos \theta_i \cos \phi_i + i \sin \phi_i,
\pm \cos \theta_i \sin \phi_i - i \cos \phi_i,
\mp \sin{\theta_i} \big)~,
\end{align}  
with $i=1,2$.

\section{The soft photon integrals  \label{sec-asoft}}
\label{appex:soft}

  The function $F(p_1,p_2,q_1,q_2,r)$ defined in~\ref{ffdef} can be split
  into three parts 
\begin{eqnarray}\label{fint}
F(p_1,p_2,q_1,q_2,r) = F_{ISR}(p_1,p_2,r) +2F_{INT}(p_1,p_2,q_1,q_2,r)
\nonumber \\
 + F_{FSR}(q_1,q_2,r) 
\end{eqnarray}
with
\begin{eqnarray}
F_{ISR}(p_1,p_2,r) = \frac{-\alpha}{4 \pi}\int \frac{d^3k_{1}}{E_{k_1}}  \left( \frac{p_1}{p_1 \cdotp  k_1} - \frac{p_2}{k_1 \cdotp p_2} \right)^2,
\end{eqnarray}

\begin{eqnarray}
F_{FSR}(q_1,q_2,r) =  \frac{-\alpha}{4 \pi}\int \frac{d^3k_{1}}{E_{k_1}}  \left( \frac{q_2}{q_2 \cdotp  k_1} - \frac{q_1}{k_1 \cdotp q_1} \right)^2,
\end{eqnarray}
and
\begin{eqnarray}
F_{INT}(p_1,p_2,q_1,q_2,r) =  
\frac{-\alpha}{4 \pi} \int \frac{d^3k_{1}}{E_{k_1}}  \left( \frac{p_1^{\mu}}{p_1 \cdotp  k_1} - \frac{p^\mu_2}{k_1 \cdotp p_2} \right)  \left( \frac{q_{2 \mu}}{q_2 \cdotp  k_1} - \frac{q_{1 \mu}}{k_1 \cdotp q_1} \right).
\end{eqnarray}

Those parts of the functions $F_{i}$ for $i=ISR,FSR,INT$, which depend explicitly
 on the ratio of the photon energy cut-off $E_{max}$ and the photon mass
  regulator $\lambda$ ($r=\frac{2E_{max}}{\lambda}$) are denoted by $F_{ir}$
 and read
\begin{eqnarray}
F_{r}(p_1,p_2,q_1,q_2,r) = F_{ISR r}(p_1,p_2,r) +2F_{INT r}(p_1,p_2,q_1,q_2,r)
\nonumber \\
 + F_{FSR r}(q_1,q_2,r)
\end{eqnarray}
with
\begin{eqnarray}
F_{ISR r}(p_1,p_2,r) = -\frac{2 \alpha}{\pi} \left[ 1-\frac{p_1p_2 \log{\left(
        \frac{ (1+\beta_{e})^4}{16m_{e}^4/s^2} \right)}}{s \beta_{e}} \right]
\log{\left( r\right)},
\end{eqnarray}
\begin{eqnarray}
F_{FSR r}(q_1,q_2,r) = -\frac{2 \alpha}{\pi} \left[ 1-\frac{q_1q_2 \log{\left( \frac{ (1+\beta_{\mu})^4}{16m_{\mu}^4/q^4} \right)}}{q^2 \beta_{\mu}} \right] \log{\left( r\right)},
\end{eqnarray}
\begin{eqnarray}
F_{INT r}(p_1,p_2,q_1,q_2,r) =  -\frac{ \alpha}{\pi} \sum^2_{i,j=1} \frac{(-1)^j p_i q_j \log{\left( \frac{ (1+\beta_{ij})^2 (p_i q_j)^2}{m_e^2 m_{\mu}^2} \right)}}{2\sqrt{(p_i q_j)^2 - m_e^2 m_{\mu}^2}} \log{\left(r\right)}, \ \
\end{eqnarray}
and
\begin{eqnarray}
\beta_i = \sqrt{1 - 4 m_i^2/s}
\nonumber \\
\beta_{ij} = \sqrt{1 - m_e^2 m_{\mu}^2/(p_i q_j)^2}.
\end{eqnarray}

  Terms proportional to $\frac{\lambda}{E_{max}}$ are neglected.
 The 'translation' to dimensional regularisation results in changing
 
 \begin{equation}
\log{\left( \frac{\lambda^2}{s}\right)} \quad
{\rm into} \quad
\Delta = \frac{(4\pi)^\epsilon}{\epsilon \Gamma(1-\epsilon)} 
\left( \frac{\mu^2}{s}\right)^\epsilon~.
\end{equation}

 The remaining parts $F_{i,fin}$ of the soft formula 
 $F_{i}=F_{ir}+F_{i,fin}$ have the following form: 
\begin{equation}
F_{ISR,fin}(q_1,q_2) = -\frac{\alpha}{4\pi^2}(I_1(p_1) + I_1(p_2) - 2I_3(p_1,p_2)),
\end{equation}

\begin{equation}
F_{FSR,fin}(q_1,q_2) = -\frac{\alpha}{4\pi^2}(I_1(q_1) + I_1(q_2) - 2I_3(q_1,q_2)),
\end{equation}

\begin{eqnarray}
F_{INT,fin}(p_1,p_2,q_1,q_2) =  -\frac{ 4\alpha}{\pi^2} \sum^2_{i,j=1} (-1)^j I_3(p_i,q_j)).
\end{eqnarray}

The arguments $x_i$ of $I_1(x_1)$ and $I_3(x_1,x_2)$ are four momenta, $x_i=(x_i(0),\bar{x_i})$:

\begin{equation}
I_1(x) = \frac{2\pi x(0)}{|\bar{x}|} \log\left(\frac{x(0) - |\bar{x}|}{x(0) + |\bar{x}|}\right),
\end{equation}

\begin{eqnarray}
I_3(x_1,x_2)) = f_1(C_{3a}I_{3a}+C_{3b}I_{3b}+C_{3c}I_{3c}+C_{3d}I_{3d}),
\end{eqnarray}

\begin{eqnarray}
f_1(x_1,x_2)\equiv f_1 = \frac{8\pi x_1x_2(|\bar{x}_2-\bar{x}_1|)^{3/2}}{|\bar{x}_1|^2\bar{x}_2|^2-(\bar{x}_1\bar{x}_2)(x_2-x_1)^2}.
\end{eqnarray}

The $I_{3a}$ function depends on the sign of $f_1$. If $f_1<0$ then:

\begin{eqnarray}\nonumber
I_{3a}(x_1,x_2)\equiv I_{3a}
&=& 
\log{\left( \frac{t_y t_3 - 1}{t_x t_3 -1} \right)} \log{\left( \frac{C_2
(t_4-t_3)t_3}{(1+t_3t_4)(1+t_3^2)} \right)}
\\ \nonumber
&&+~\frac{1}{2} \log^2{\left( \frac{t_y t_3 - 1}{t_3} \right)}-\frac{1}{2} \log^2{\left( \frac{t_x t_3 -
1}{t_3} \right)} - \textrm{Li$_2$}\left( \frac{(t_yt_3-1)t_4}{t_3-t_4} \right)
\\ \nonumber
&&+~ \textrm{Li$_2$}\left( \frac{(t_xt_3-1)t_4}{t_3-t_4} \right) + \textrm{Li$_2$}\left(
\frac{1-t_3t_y}{1+t_3^2} \right)-\textrm{Li$_2$}\left( \frac{1-t_3t_x}{1+t_3^2} \right)
\\ 
&&+~ \textrm{Li$_2$}\left( \frac{1-t_3t_y}{1+t_3t_4} \right) - \textrm{Li$_2$}\left( \frac{1-t_3t_x}{1+t_3t_4}
\right)
\end{eqnarray}If $f_1>0$ then: 

\begin{eqnarray}\nonumber
I_{3a}(x_1,x_2)\equiv I_{3a} = \log{\left( \frac{t_y t_3 - 1}{t_x t_3 -1} \right)} \log{\left( \frac{C_2 (t_4-t_3)t_3}{(1+t_3^2)} \right)}
\\ \nonumber
+\frac{1}{2} \log^2{\left( \frac{1-t_y t_3}{t_3} \right)}-\frac{1}{2} \log^2{\left( \frac{1-t_x t_3}{t_3} \right)} - \textrm{Li$_2$}\left( \frac{(t_yt_3-1)t_4}{t_3-t_4} \right)
\\ \nonumber
+ \log{\left( \frac{1+t_3 t_4 - 1}{ t_3} \right)}\log{\left( \frac{t_y +t_4}{t_x +t_4} \right)} - \log{\left( \frac{1-t_y t_3}{ t_3} \right)}\log{\left( -t_y-t_4 \right)}
\\ \nonumber
 + \log{\left( \frac{1-t_x t_3}{ t_3} \right)}\log{\left( -t_x-t_4 \right)}
+ \textrm{Li$_2$}\left( \frac{(t_xt_3-1)t_4}{t_3-t_4} \right)
\\ \nonumber
 + \textrm{Li$_2$}\left( \frac{1-t_3t_y}{1+t_3^2} \right)
-\textrm{Li$_2$}\left( \frac{1-t_3t_x}{1+t_3^2} \right)
- \textrm{Li$_2$}\left( \frac{(t_y-t_4)t_3}{1+t_3t_4} \right)
\\ 
+\textrm{Li$_2$}\left( \frac{(t_x-t_4)t_3}{1+t_3t_4} \right)
\end{eqnarray}

\begin{eqnarray}\nonumber
I_{3b}(x_1,x_2)\equiv I_{3b} = \log{\left( \frac{t_y t_4 - 1}{t_x t_4 -1} \right)} \log{\left( \frac{C_2 (t_3-t_4)}{(1+t_4^2)t_3} \right)} + \frac{1}{2} \log^2{\left( \frac{t_y t_4 - 1}{t_4} \right)}
\\ \nonumber - \frac{1}{2} \log^2{\left( \frac{t_x t_4 - 1}{t_4} \right)} - \log{\left( \frac{t_y t_4 - 1}{t_4} \right)} \log{\left( t_y-t_3 \right)}
\\ \nonumber
 + \log{\left( \frac{t_x t_4 - 1}{t_4} \right)} \log{\left( t_x-t_3 \right)} + 
\log{\left( \frac{-t_3 t_4 - 1}{t_4} \right)} \log{\left( \frac{t_y+t_3}{t_x+t_3} \right)}
\\ \nonumber
-\textrm{Li$_2$}\left( \frac{(1-t_yt_4)t_4}{t_3-t_4} \right)+\textrm{Li$_2$}\left( \frac{(1-t_xt_4)t_4}{t_3-t_4} \right) -\textrm{Li$_2$}\left( \frac{(t_y+t_3)t_4}{1+t_3t_4} \right)
\\ 
+\textrm{Li$_2$}\left( \frac{(t_x+t_3)t_4}{1+t_3t_4} \right) + \textrm{Li$_2$}\left( \frac{1-t_yt_4}{1+t_4^2} \right)-\textrm{Li$_2$}\left( \frac{1-t_xt_4}{1+t_4^2} \right)  \ \
\end{eqnarray}

\begin{eqnarray}\nonumber
I_{3c}(x_1,x_2)\equiv I_{3c} = \log{\left( \frac{t_y+ t_3}{t_x+t_3} \right)} \log{\left( \frac{C_2 (1+t_3^2)(1+t_3t_4}{t_3t_4(t_4-t_3)} \right)}
\\ \nonumber
 - \frac{1}{2} \log^2{\left( t_y+t_3 \right)}+ \frac{1}{2} \log^2{\left( t_x+t_3 \right)}-\textrm{Li$_2$}\left( \frac{(t_y+t_3)t_3}{1+t_3^2} \right)
\\ \nonumber
+ \textrm{Li$_2$}\left( \frac{(t_x+t_3)t_3}{1+t_3^2} \right) -\textrm{Li$_2$}\left( \frac{(t_y+t_3)t_4}{1+t_3t_4} \right) + \textrm{Li$_2$}\left( \frac{(t_x+t_3)t_4}{1+t_3t_4} \right)
\\
+ \textrm{Li$_2$}\left( \frac{(t_y+t_3)t_4}{t_3 - t_4} \right) - \textrm{Li$_2$}\left( \frac{(t_x+t_3)t_4}{t_3 - t_4} \right)
\end{eqnarray}

\begin{eqnarray}\nonumber
I_{3d}(x_1,x_2)\equiv I_{3d}( = - \log{\left( \frac{t_y+ t_4}{t_x+t_4} \right)} \log{\left( \frac{C_2 (1+t_4^2)(1+t_3t_4}{t_3t_4(t_4-t_3)} \right)}
\\ \nonumber
+ \frac{1}{2} \log^2{\left( -t_y-t_4 \right)} - \frac{1}{2} \log^2{\left( -t_x-t_4 \right)}  + \textrm{Li$_2$}\left( \frac{(t_y+t_4)t_3}{1+t_3t_4} \right)
\\ \nonumber
- \textrm{Li$_2$}\left( \frac{(t_x+t_4)t_3}{1+t_3t_4} \right)+\textrm{Li$_2$}\left( \frac{(t_y+t_4)t_4}{1+t_4^2} \right)-\textrm{Li$_2$}\left( \frac{(t_x+t_4)t_4}{1+t_4^2} \right)
\\ 
-\textrm{Li$_2$}\left( \frac{t_y+t_4}{t_4-t_3} \right)+\textrm{Li$_2$}\left( \frac{t_x+t_4}{t_4-t_3} \right)
\end{eqnarray}

\begin{eqnarray}\nonumber
C_{3a}(x_1,x_2)\equiv C_{3a} = \frac{2(x_2(0)-x_1(0))\sqrt{|\bar{x}_1|^2|\bar{x}_2|^2 - \bar{x}_1\bar{x}_2}(t_3^2-1)}{(|\bar{x}_1|^2+|\bar{x}_2|^2 - 2\bar{x}_1\bar{x}_2)t_3^2(t_3+\frac{1}{t_3})(t_4+\frac{1}{t_3})(\frac{1}{t_3}-\frac{1}{t_4})}
\nonumber \\
+\frac{4(x_2(0)-x_1(0))(|\bar{x}_1|^2-\bar{x}_1\bar{x}_2)t_3}{(|\bar{x}_1|^2+|\bar{x}_2|^2 - 2\bar{x}_1\bar{x}_2)t_3^2(t_3+\frac{1}{t_3})(t_4+\frac{1}{t_3})(\frac{1}{t_3}-\frac{1}{t_4})}
\nonumber \\
+\frac{4x_1(0)}{t_3(t_3+\frac{1}{t_3})(t_4+\frac{1}{t_3})(\frac{1}{t_3}-\frac{1}{t_4})} \ \
\end{eqnarray}

\begin{eqnarray}\nonumber
C_{3b}(x_1,x_2)\equiv C_{3b} = -\frac{2(x_2(0)-x_1(0))\sqrt{|\bar{x}_1|^2|\bar{x}_2|^2 - \bar{x}_1\bar{x}_2}(t_4^2-1)}{(|\bar{x}_1|^2+|\bar{x}_2|^2 - 2\bar{x}_1\bar{x}_2)t_4^2(\frac{1}{t_3}-\frac{1}{t_4})(t_3+\frac{1}{t_4})(t_4+\frac{1}{t_4})}
\\ \nonumber
-\frac{4(x_2(0)-x_1(0))(|\bar{x}_1|^2-\bar{x}_1\bar{x}_2)t_3}{(|\bar{x}_1|^2+|\bar{x}_2|^2 - 2\bar{x}_1\bar{x}_2)t_4^2(\frac{1}{t_3}-\frac{1}{t_4})(t_3+\frac{1}{t_4})(t_4+\frac{1}{t_4})}
\\ 
-\frac{4x_1(0)}{t_4(\frac{1}{t_3}-\frac{1}{t_4})(t_3+\frac{1}{t_4})(t_4+\frac{1}{t_4})} \ \
\end{eqnarray}

\begin{eqnarray}\nonumber
C_{3c}(x_1,x_2)\equiv C_{3c} = \frac{2(x_2(0)-x_1(0))\sqrt{|\bar{x}_1|^2|\bar{x}_2|^2 - \bar{x}_1\bar{x}_2}(t_3^2-1)}{(|\bar{x}_1|^2+|\bar{x}_2|^2 - 2\bar{x}_1\bar{x}_2)(t_3+\frac{1}{t_3})(t_3+\frac{1}{t_4})(t_3-t_4)}
\\ \nonumber
+\frac{4(x_2(0)-x_1(0))(|\bar{x}_1|^2-\bar{x}_1\bar{x}_2)t_3}{(|\bar{x}_1|^2+|\bar{x}_2|^2 - 2\bar{x}_1\bar{x}_2)(t_3+\frac{1}{t_3})(t_3+\frac{1}{t_4})(t_3-t_4)}
\\ 
+\frac{4x_1(0)t_3}{(t_3+\frac{1}{t_3})(t_3+\frac{1}{t_4})(t_3-t_4)} \ \
\end{eqnarray}

\begin{eqnarray}\nonumber
C_{3d}(x_1,x_2)\equiv C_{3d} = \frac{2(x_2(0)-x_1(0))\sqrt{|\bar{x}_1|^2|\bar{x}_2|^2 - \bar{x}_1\bar{x}_2}(1-t_4^2)}{(|\bar{x}_1|^2+|\bar{x}_2|^2 - 2\bar{x}_1\bar{x}_2)(t_4+\frac{1}{t_3})(t_4+\frac{1}{t_4})(t_3-t_4)}
\\ \nonumber
-\frac{4(x_2(0)-x_1(0))(|\bar{x}_1|^2-\bar{x}_1\bar{x}_2)t_4}{(|\bar{x}_1|^2+|\bar{x}_2|^2 - 2\bar{x}_1\bar{x}_2)(t_4+\frac{1}{t_3})(t_4+\frac{1}{t_4})(t_3-t_4)}
\\ 
-\frac{4x_1(0)t_4}{(t_4+\frac{1}{t_3})(t_4+\frac{1}{t_4})(t_3-t_4)} \ \
\end{eqnarray}

\begin{eqnarray}
t_x(x_1,x_2)\equiv t_x = \frac{|\bar{x}_1|^2-\bar{x}_1\bar{x}_2+|\bar{x}_1||\bar{x}_1-\bar{x}_2|}{\sqrt{|\bar{x}_1|^2|\bar{x}_1|^2 - (\bar{x}_1\bar{x}_2)^2}}
\end{eqnarray}

\begin{eqnarray}
t_y(x_1,x_2)\equiv t_y = \frac{2\bar{x}_1\bar{x}_2 - |\bar{x}_1|^2-|\bar{x}_2|^2+|\bar{x}_2||\bar{x}_1+\bar{x}_2|}{\sqrt{|\bar{x}_1|^2|\bar{x}_1|^2 - (\bar{x}_1\bar{x}_2)^2}}
\end{eqnarray}

\begin{eqnarray}
t_{3,4}(x_1,x_2)\equiv t_{3,4} =  \frac{x_1(0)(|\bar{x}_2|^2-\bar{x}_1\bar{x}_2)+x_2(0)(|\bar{x}_1|^2-\bar{x}_1\bar{x}_2)\mp\sqrt{\Delta}}{\sqrt{|\bar{x}_1|^2|\bar{x}_2|^2-(\bar{x}_1\bar{x}_2)^2}((x_2(0)-x_1(0))+|\bar{x}_2-\bar{x}_1|)}
\end{eqnarray}

\begin{eqnarray}
\Delta = (x_1(0)(|\bar{x}_2|^2-\bar{x}_1\bar{x}_2)+x_2(0)(|\bar{x}_1|^2-\bar{x}_1\bar{x}_2))^2
\\ \nonumber + (|\bar{x}_1|^2|\bar{x}_2|^2-(\bar{x}_1\bar{x}_2)^2)(x_2-x_1)^2
\end{eqnarray}

\begin{eqnarray}
C_2 = \frac{(x_2-x_1)^2}{(x_2(0)-x_1(0)+|\bar{x}_2-\bar{x}_1|)^2}.
\end{eqnarray}

\providecommand{\href}[2]{#2}
\addcontentsline{toc}{section}{References} 
\bibliography{mu_nlo}

\end{document}